\documentclass[12pt,notoc]{JHEP3}

\usepackage{amsmath,amssymb,euscript,array,cite,mathrsfs}

\usepackage{epsfig}

\def\a{\alpha}
\def\b{\beta}
\def\c{\gamma}
\def\d{\delta}
\def\e{\epsilon}

\def\k{\kappa}
\def\l{\lambda}
\def\m{\mu}
\def\n{\nu}

\def\r{\rho}
\def\s{\sigma}
\def\t{\tau}
\def\u{\upsilon}
\def\w{\omega}

\def\D{\Delta}
\def\L{\Lambda}

\def\S{\Sigma}
\def\W{\Omega}

\def\vare{\varepsilon}

\def\sst{\scriptscriptstyle}
\def\det{{\rm det}}

\def\BcalF{\boldsymbol{\cal F}}

\def\BE{\boldsymbol{E}}
\def\Bvartheta{\boldsymbol{\vartheta}}
\def\Bone{\boldsymbol{1}}

\def\R{\boldsymbol{R}}

\def\BDelta{\boldsymbol{\Delta}}

\def\BA{\boldsymbol{A}}
\def\BB{\boldsymbol{B}}
\def\BC{\boldsymbol{C}}
\def\BK{\boldsymbol{K}}
\def\Bn{\boldsymbol{n}}
\def\Bs{\boldsymbol{s}}
\def\Bh{\boldsymbol{h}}
\def\BC{\boldsymbol{C}}
\def\BOmega{\boldsymbol{\Omega}}
\def\orta{\overrightarrow}

\newcommand{\SLASH}[1]{{\raise.15ex\hbox{/}\mkern-12mu #1}}
\def\Dbarslash{\,\,{\raise.15ex\hbox{/}\mkern-12mu {\bar D}}}
\def\Dslash{\,\,{\raise.15ex\hbox{/}\mkern-12mu D}}
\def\delslash{\,\,{\raise.15ex\hbox{/}\mkern-9mu \partial}}
\def\delbarslash{\,\,{\raise.15ex\hbox{/}\mkern-9mu {\bar\partial}}}

\def\rta{\rightarrow}
\newcommand{\IM}{\operatorname{Im}}
\newcommand{\RE}{\operatorname{Re}}

\newcommand{\MAT}[1]{\begin{pmatrix} #1\end{pmatrix}}
\newcommand{\EQ}[1]{\begin{equation} #1 \end{equation}}

\newcommand{\SP}[1]{\begin{equation}\begin{split} #1
\end{split}\end{equation}}

\title{The Causal Structure of QED in Curved Spacetime:~
Analyticity and the Refractive Index}
\author{Timothy J. Hollowood and Graham M. Shore\\
Department of Physics,\\ University of Wales Swansea,\\
Swansea, SA2 8PP, UK.\\
E-mail: {\tt t.hollowood@swansea.ac.uk, g.m.shore@swansea.ac.uk}}
\abstract{
The effect of vacuum polarization on the propagation of photons in
curved spacetime is studied in scalar QED. A compact formula is given
for the full frequency dependence of the refractive index for any
background in terms of the Van Vleck-Morette matrix for its Penrose
limit and it is shown how the superluminal propagation found in the 
low-energy effective action is reconciled with causality.
The geometry of null geodesic congruences is found to imply a novel
analytic structure for the refractive index and Green functions of
QED in curved spacetime, which preserves their causal nature but
violates familiar axioms of $S$-matrix theory and dispersion relations. 
The general formalism is illustrated in a number of examples,
in some of which it is found that the refractive index develops
a negative imaginary part, implying an amplification of photons 
as an electromagnetic wave propagates through curved spacetime.}

\begin{document}

\section{Introduction}

Quantum field theory in curved spacetime has proved to be a rich field
exhibiting many subtle and counter-intuitive phenomena. The most famous,
of course, is the prediction of Hawking radiation from black holes 
\cite{Hawking:1974sw},
which has forced a critical analysis of unitarity in spacetimes with
horizons. More recently, the insights associated with holography 
\cite{tHooft:1993gx,Susskind:1994vu} 
have led to a re-appraisal of the r\^ole of locality at a fundamental 
level. Another remarkable, but less well-known, phenomenon discovered
during the early investigations of QFT in curved spacetime is the
apparent superluminal propagation of photons due to vacuum
polarization in QED.  
This clearly raises the question of whether causality may be violated 
by quantum effects in curved spacetime. 

The original result, due to Drummond and Hathrell \cite{Drummond:1979pp}, 
was obtained by constructing the effective action for QED in a curved 
background and shows that the low-frequency limit of the phase velocity 
$v_{ph}(\omega) = c/n(\omega)$, where $n(\omega)$ is the refractive index, 
can exceed the fundamental speed-of-light constant $c$. 
This is not immediately paradoxical\footnote{For a review of the issues
involved in reconciling superluminal propagation with causality for
QED in curved spacetime, see 
refs.~\cite{Shore:1995fz,Shore:2003jx,Shore:2003zc,Shore:2007um}.
Related work on causality can be found, 
{\it e.g.}, in refs.~\cite{Dolgov:1997hc,Liberati:2001sd,Dubovsky:2007ac}.} 
since the `speed of light' relevant 
for causality is not $v_{ph}(0)$ but the wavefront velocity, 
which can be identified with the high-frequency limit 
$v_{ph}(\infty)$ \cite{Leontovich}. In order to settle the question of
causality, it is therefore necessary to go beyond the low-energy  
effective action and show explicitly that $n(\infty) = 1$.
However, a serious problem then arises because of the Kramers-Kronig 
dispersion relation \cite{Kramers,Kronig,Weinberg}, 
which is proved in Minkowski spacetime on the basis of apparently
fundamental axioms, especially micro-causality, together with standard
analyticity properties of QFT amplitudes. This states: 
\EQ{
n(0) - n(\infty) =  \frac{2}{\pi} \int_0^\infty d\omega\,
\text{Im}\,n(\omega)\qquad\qquad(\text{Minkowski})
\label{wqw1}
}
Since unitarity, in the form of the optical theorem, normally implies
that $\text{Im}\,n(\omega)$ is positive, a superluminal $n(0) < 1$
would seem to imply a superluminal wavefront velocity, $n(\infty) <
1$, with the associated violation of causality. 

The resolution of this apparent paradox was found in our recent papers
\cite{Hollowood:2007ku,Hollowood:2007kt}. 
We showed there that generic geometrical properties of null geodesics 
in curved spacetime imply a novel analytic structure for the refractive
index which invalidates the Kramers-Kronig relation, at least in the form 
\eqref{wqw1}. The complete frequency dependence of the 
refractive index was found in simple examples and it was shown explicitly 
how a superluminal $n(0)$ is reconciled with $n(\infty) = 1$, ensuring
causality.  
 
An important implication of this result is that the conventional
assumptions about the analytic structure of amplitudes in QFT, which
underpin the whole of $S$-matrix theory and dispersion relations, have
to be reassessed in curved spacetime. Because of the intimate relation
of analyticity and causality, this is key issue both for QFT in curved
spacetime and, most likely, for quantum gravity itself. It also
highlights the danger in theories involving gravity of relying on
identities and intuition derived from conventional dispersion
relations to extrapolate from low-energy effective field theories to
their UV completions. In particular, the occurrence of `superluminal'
behaviour in a low-energy theory does not necessarily mean that such
theories do not have consistent, causal UV completions, either in QFT
or string theory 
\cite{Adams:2006sv,Distler:2006if,Bonvin:2006vc,Babichev:2007dw}. 
 
The centrepiece of the present paper is the derivation of a formula for the 
full frequency dependence of the refractive index for QED in an arbitrary 
curved spacetime expressed entirely geometrically, specifically in terms of 
the Van Vleck-Morette (VVM) matrix in the Penrose limit
\cite{Penrose,Blau2,Blau:2006ar}.  The calculation 
uses conventional QED Feynman diagram methods with the heat kernel/proper 
time formulation of the propagators, rather than the worldline method used 
in our earlier papers. This allows us to retain the critical insight of the 
worldline approach in motivating the importance of the Penrose limit, while 
strengthening the contact with the well-developed differential geometry of 
null geodesic congruences in general relativity.

This geometry plays a central r\^ole here both in the derivation of our 
key formula for the refractive index and in the interpretation of its 
analytic structure. In particular, the idea of the Penrose limit is 
vital in establishing the generality of our results. The fundamental 
insight provided by the worldline analysis is that to leading order in 
$R\lambda_c^2$ (where $R$ is a typical curvature and 
$\lambda_c = 1/m$, the Compton wavelength of the electron, sets the
quantum scale), the one-loop corrections to photon propagation are
governed by  fluctuations around the null geodesic describing the
classical photon trajectory. It is precisely this geometry of geodesic
deviation in the original curved spacetime that is encoded in the
Penrose plane-wave limit. This also explains why the final result for
the refractive index can be expressed purely in terms of the VVM
matrix since, as we explore here in some detail, this is in turn
determined by the Jacobi fields characterizing geodesic deviation
\cite{HawkingEllis,Visser:1992pz,Wald:1984rg}. 

A crucial feature of the geometry of null geodesic congruences is the 
occurrence of conjugate points, {\it i.e.}~two points on a null geodesic 
which can be joined by an infinitesimal deformation of the original 
geodesic \cite{HawkingEllis,Wald:1984rg}. 
Their occurrence is generic given the validity of the null energy
condition, which is an important assumption in most theorems involving
causality, horizons and singularities in general relativity. The
existence of conjugate points implies singularities at the
corresponding points in the VVM matrix. Translated into the quantum
field theory, these imply singularities in the refractive index in the
complex  
$\omega$-plane -- in particular, $n(\omega)$ must be defined on a physical 
sheet with cuts running on the real axis from 0 to $\pm \infty$.
This novel analytic structure has important consequences, most notably the 
loss of the fundamental $S$-matrix property of real analyticity for the 
refractive index, {\it i.e.}~$n(\omega^*) = n(\omega)^*$, which is 
assumed in the derivation of the Kramers-Kronig relation \eqref{wqw1}. 
The relation with causality means that analyticity is a key property of 
QFT amplitudes, so we must show how, despite the violation of the 
Kramers-Kronig relation, the new analytic structure of the refractive 
index is reconciled with, and indeed essential for, causality.

It is important to emphasize that the essential physics underlying this 
discussion is much more general than the specific application to the 
refractive index in QED. It shows how the geometry of curved spacetime 
can modify the analytic structure of Green functions and scattering 
amplitudes in quantum field theory in a quite radical way. This is sure 
to have important physical consequences which we have only just begun 
to explore. Certainly, the implications for $S$-matrix theory and 
dispersion relations appear to be far-reaching.

Of course, a full discussion of causality, and micro-causality, must 
be framed more generally in terms of the Green functions of the theory.
In this paper, we explicitly construct the one-loop corrected Green
functions for QED in the Penrose plane-wave spacetime, which is
sufficient to address the issues of causality in photon propagation.
The full range of Green functions---Feynman, Wightman, retarded and
advanced, commutator (Pauli-Jordan or Schwinger)---is found and they
are 
shown to exhibit the expected good causality properties. In particular,
the retarded (advanced) Green functions are shown to have support only
on or inside the forward (backward) light cone. This confirms that,
even at one-loop, the commutator function vanishes outside the light cone,
which is the conventional quantum field theoretic definition of 
micro-causality.

\vskip0.2cm
The paper is organized as follows. We begin with a review of the 
classical theory of wave propagation in curved spacetime in the eikonal
approximation. We then summarize the low-energy effective field theory, 
extending the Drummond-Hathrell result to scalar QED.\footnote{The case
of spinor QED is similar in terms of physics to the results presented here.
However, the formalism requires further technical developments and will 
be presented separately.}
Our main result is presented in Section 4, where we calculate the one-loop 
vacuum polarization in the Penrose plane-wave limit and derive the 
fundamental formula for the refractive index in terms of the VVM matrix.
Section 5 reviews the geometry of geodesic deviation and a number of 
important identities relating the VVM matrix, geodesic interval and 
Jacobi fields are derived. The Raychoudhuri equations are used to
demonstrate the generic nature of conjugate points.  

The analytic structure of the refractive index is studied in Section 6. 
The argument leading from the existence 
of singularities in the VVM matrix to the definition of the physical 
sheet for the refractive index in the complex $\omega$-plane is
explored and the consequences for the Kramers-Kronig relation and
causality are carefully discussed. The explicit construction of the
retarded, advanced 
and commutator Green functions at one-loop, demonstrating that they have
the required causal properties, is presented in Section 9.

These formal results are illustrated in Sections 7 and 8 in a number of
examples, demonstrating explicitly the predicted relation of the geometry
with analyticity and causality. The cases of conformally flat and
Ricci flat symmetric plane waves, and also weak gravitational waves,
are calculated in detail, with the latter two exhibiting gravitational
birefringence. Remarkably, we also find that the refractive index may
develop a negative imaginary part, contrary to the conventional  
flat-spacetime expectation based on unitarity and the optical
theorem. Although the physical origin of this effect remains to be
fully understood, it corresponds to a quantum mechanical amplification
of the electromagnetic wave as it passes through the curved background
spacetime, over and above 
the geometric effects of focusing or defocusing, apparently due to
the emission of photons induced by the interaction with the
background field.  Finally, our conclusions are summarized in Section
10.

\section{Classical Photon Propagation in Curved Spacetime}

The classical propagation of photons in curved spacetime 
is governed by the covariant Maxwell equation,
\EQ{
\nabla_\mu F^{\mu\nu}=0\ ,\qquad F_{\mu\nu}=\nabla_\mu A_\nu-
\nabla_\nu A_\mu\ .
}
In a general background spacetime, it is not possible to solve these
equations exactly. However, we will 
work in the eikonal, or WKB, approximation which is valid when the
frequency is much greater than the scale over which the 
curvature varies, $\omega\gg\sqrt
R$. (Here, $R$ is a measure of the curvature scale, for instance a
typical element of the Riemann tensor.)  
In this case, we can write the electromagnetic field in the form
\EQ{
A_\mu(x)=\varepsilon_\mu(x)e^{-i\Theta(x)}\ ,
\label{wkb}
}
where the eikonal phase $\Theta$ is ${\cal O}(\omega)$ and 
$\varepsilon_\mu$ is ${\cal O}(\omega^0)$. Substituting into Maxwell's
equation, and expanding in powers of $1/\omega$, we find the the
leading and next-to-leading order terms are
\EQ{
\nabla_\mu F^{\mu\nu}=\Big[-
\partial\Theta\cdot\partial\Theta
\varepsilon^\nu +
2i\partial\Theta\cdot\nabla\,\varepsilon^\nu+i\varepsilon^\nu
\nabla\cdot\partial\Theta+\cdots\Big]e^{-i\Theta}\ .
\label{xcc}
}
The leading order term yields the {\it eikonal equation\/}
\EQ{
\partial\Theta\cdot\partial\Theta=0\ ;
\label{ngr}
}
so the gradient $k^\mu=\partial^\mu\Theta$ is a null vector
field. This vector field defines a {\it null congruence\/}, that is a
family of null geodesics whose tangent vectors are identified with the
vector field $k^\mu$. This vector can also be identified with the 
4-momentum of photons: in this sense the eikonal approximation is the
limit of classical ray optics.

It is convenient to introduce
a set of coordinates  $(u,V,Y^a)$, $a=1,2$, the {\it Rosen coordinates\/},
that are specifically adapted to the null congruence: $u$ is the
affine parameter along the geodesics; $V$ is the associated null
coordinate so that 
\EQ{
\Theta=\omega V\ ;
}
while $Y^a$ are two orthogonal space-like coordinates.
As explained in ref.~\cite{Blau2}, the full metric
$g_{\mu\nu}$ can always be brought into the form
\EQ{
ds^2=-2du\, d V+C(u, V,Y^a)dV^2+2C_a(u, V,Y^b)
dY^a\,d V+C_{ab}(u, V,Y^c)dY^a\,dY^b\ .
\label{mrosen}
}
The null congruence
has a simple description as the set of curves  $(u,V,Y^a)$ for fixed 
values of the transverse coordinates $(V,Y^a)$. 
It should not be surprising that the
Rosen coordinates are singular at the {\it caustics\/} of the
congruence, that is points where members of the congruence
intersect.

The next-to-leading order in the eikonal approximation \eqref{xcc} gives
an equation for the evolution of $\varepsilon^\mu$ along a null geodesic:
\EQ{
k\cdot \nabla\,\varepsilon^\mu=-\tfrac12\varepsilon^\mu\nabla\cdot k\ .
\label{pip}
}
It is useful to make the decomposition $\varepsilon^\mu={\cal A}
\hat\varepsilon^\mu$, where $\hat\varepsilon^\mu$ is the unit
normalized polarization vector and ${\cal A}$ represents the amplitude.
Eq.\eqref{pip} is then equivalent to the two equations
\SP{
k\cdot\nabla\,\hat\varepsilon^\mu&=0\ ,\\
k\cdot\nabla\,\log{\cal A}&=-\tfrac12\nabla\cdot k\ .
\label{pippip}
}
At this point, we fix the gauge by choosing $A^u=0$ along with the
condition $\nabla_\mu A^\mu=0$. The latter implies the 
transverse condition $k\cdot\hat\varepsilon=0$ whilst the former means
that we set the component of $\hat\varepsilon$ along $k$ to
zero. Hence, there are two independent solutions for the polarization vector 
$\hat\varepsilon_{(i)}$, $i=1,2$, which we normalize as 
$\hat\varepsilon_{(i)}\cdot\hat\varepsilon_{(j)}=\delta_{ij}$. 
These span the directions associated to the space-like coordinates $Y^a$.
The second of eqs.\eqref{pippip} relates the change of the amplitude along 
a null geodesic to the {\it expansion} $\hat\theta = \nabla_\mu k^\mu$,
one of the optical scalars appearing in the Raychoudhuri equations
(see Section 5).

Later, when we calculate the one-loop correction to the mass-shell
condition we shall have to take the photon wavefunctions
off-shell at tree level. This can be done conveniently by modifying the eikonal
phase to
\EQ{
\Theta=\omega\big(V-\vartheta_{ij}(u;\omega)\big)\ .
\label{cor}
}
We have indicated the polarization dependence explicitly, in which
case the phase can be thought of as a $2\times2$ matrix with
\EQ{
A^\mu_{(i)}=\varepsilon^\mu_{(j)}
e^{-i\omega(V-\vartheta_{ij}(u;\omega))}\ ,
\label{dds}
}
with an implicit sum over $j=1,2$, 
in which case, 
\EQ{
\nabla_\mu
F_{(i)}^{\mu\nu}=2\omega^2\frac{\partial
\vartheta_{ij}(u;\omega)}{\partial u}\varepsilon_{(j)}^\nu
e^{-i\omega V}\ ,
\label{os}
}
to leading order in the eikonal approximation. 
If $\vartheta_{ij}(u;\omega)$ is perturbatively small then the local 
phase velocity matrix 
is $c(\delta_{ij}-\partial\vartheta_{ij}(u;\omega)/\partial u)$, which gives a
matrix of refractive indices 
\EQ{
\Bn(u;\omega)=\Bone+2\frac{\partial\Bvartheta(u;\omega)}{\partial u}\ .
\label{rfr}
}
Notice that in order for the correction to remain perturbatively
small, the refractive index should strictly-speaking 
approach $\Bone$ in the infinite past and future. In  other words the
spacetime should become flat in these limits.

\section{Effective Action and Low-Frequency Propagation}

The low-frequency limit of the phase velocity, which exhibits the superluminal
effect, can be found by considering the modifications to the Maxwell equation
following from the leading terms in a derivative expansion of the one-loop effective action. This was the approach taken in the original work of
Drummond and Hathrell \cite{Drummond:1979pp}. 

The generalization of the QED effective action to all orders in derivatives
was subsequently given in ref.\cite{Shore:2002gw,Shore:2002gn}, 
extracting the relevant ``$RFF$'' terms from the general heat kernel
results of Barvinsky {\it et al.\/} \cite{BGVZone}. 
(See also refs.~\cite{Gilkey:1975iq,BGVZtwo,Avramidi:1997jy,Barvinsky:2003rx}
for related heat kernel results.) Although these
results were given for spinor QED, it is straightforward to find the
corresponding results for scalar QED from the formulae in 
\cite{Shore:2002gw}.
In particular, this allows us to write the leading-order effective action
for scalar QED and deduce the corresponding low-frequency phase velocity,
providing a useful consistency check on our general result for the
full refractive index in scalar QED. 

The relevant terms in the effective action to one loop are
\SP{
\Gamma&=\int d^4x\,\sqrt{g}
\Big[-\tfrac14F_{\mu\nu}F^{\mu\nu}+aRF_{\mu\nu}F^{\mu\nu}+bR_{\mu\nu}
F^{\mu\lambda}F^\nu{}_\lambda\\ 
&\qquad\qquad+c
R_{\mu\nu\lambda\rho}F^{\mu\nu}F^{\lambda\rho}
+d\nabla_\mu F^{\mu\lambda}\nabla_\nu
F^\nu{}_\lambda+\cdots\Big]
\label{dhaction}
}
where, in the notation of \cite{Shore:2002gw}, 
\SP{
a&=-\frac{\alpha{\mathscr N}}{8\pi m^2}h_1(0)\ ,\\
b&=-\frac{\alpha{\mathscr N}}{8\pi m^2}\big(h_2(0)-2h_0'(0)\big)\ ,\\
c&=-\frac{\alpha{\mathscr N}}{8\pi m^2}\big(h_3(0)+h_0'(0)\big)\ ,\\
d&=-\frac{\alpha{\mathscr N}}{8\pi m^2}2h_0'(0)\ .
}
For scalar QED ${\mathscr N}=1$ while for spinor QED ${\mathscr N}=2^2=4$.
Notice that we have used the identity
\EQ{
\int d^4x\,\sqrt{g}F_{\mu\nu}\square F^{\mu\nu}
=\int d^4x\,\sqrt{g}\Big[2\nabla_\mu F^{\mu\lambda}\nabla_\nu
F^\nu{}_\lambda-2R_{\mu\nu}
F^{\mu\lambda}
F^\nu{}_\lambda-R_{\mu\nu\lambda\rho}F^{\mu\nu}F^{\lambda\rho}\Big]
}
to write the action in the form \eqref{dhaction}.

The all-orders effective action derived in \cite{Shore:2002gw} is
expressed in terms of $RFF$-type operators acted on by functions of
the Laplacian given 
in terms of the form factors $f(\square)$ and 
$F(\square_1,\square_2,\square_3)$ computed in ref.\cite{BGVZone}. 
In particular, the quantities $h(0)$ can be expressed in terms of this
collection of form factors as follows: 
\SP{
h_0'(0)&=-\tfrac12f'_4(0)+f'_5(0)\ ,\\
h_1(0)&=\tfrac18F_1(\underline{0})-\tfrac1{12}F_3(\underline{0})\ ,\\
h_2(0)&=\tilde F_8(\underline{0})\ ,\\
h_3(0)&=-\tfrac12F_3(\underline{0})+\tilde F_0(\underline{0})\ .
}
For spinor QED \cite{Shore:2002gw},
\SP{
&f'_4(0)=-\tfrac1{12}\ ,\qquad f'_5(0)=-\tfrac1{120}\ ,
\qquad F_1(\underline{0})=\tfrac16\ ,\\ &\tilde
F_8(\underline{0})=-\tfrac1{180}\ ,\qquad
F_3(\underline{0})=\tfrac1{12}\ ,\qquad 
\tilde F_0(\underline{0})=\tfrac1{72}\ .
}
So the coefficients are 
\EQ{
a = -\frac{\alpha}{144\pi m^2}\ , ~~~~~~
b = \frac{13\alpha}{360\pi m^2}\ , ~~~~~~
c = -\frac{\alpha}{360\pi m^2}\ , ~~~~~~
d = -\frac{\alpha}{30\pi m^2} \ .
}
reproducing the original Drummond-Hathrell effective action 
\cite{Drummond:1979pp}.

For scalar QED, we can readily see that $f_4=0$, $F_1=0$ and $F_3=0$, 
while the other quantities are as above, so the coefficients in this
case are
\EQ{
a = 0\ , ~~~~~~
b = \frac{\alpha}{720\pi m^2}\ , ~~~~~~
c = \frac{\alpha}{1440\pi m^2}\ , ~~~~~~
d = -\frac{\alpha}{480\pi m^2} \ .
}

To relate the effective action to the calculation of the refractive index 
we write the modified Maxwell equation corresponding to \eqref{dhaction}
and substitute the eikonal ansatz \eqref{dds}. This gives the general
result for the low-frequency limit of the refractive index
\cite{Drummond:1979pp,Shore:2003zc}
\EQ{
n_{ij}(x;0)=\delta_{ij}-2b R_{uu}(x)\delta_{ij}-8cR_{uiuj}(x)\ .
}
Notice that the refractive index, and the phase velocity, is a local
quantity in spacetime.
Although we call this the ``low-frequency'' limit, we are still
working in the eikonal approximation. Low frequency refers to the fact
that the dimensionless ratio $\omega\sqrt R/m^2$ is small. 
With this in mind, for spinor QED we find
\EQ{
n_{ij}(x;0)=\delta_{ij}-\frac{\alpha}{180\pi
  m^2}\big(13R_{uu}(x)\d_{ij}-4R_{uiuj}(x)\big)\ ,
\label{nzerospin}
}
while for scalar QED,
\EQ{
n_{ij}(x;0)=\delta_{ij}-\frac{\alpha}{360\pi
  m^2}\big(R_{uu}(x)\d_{ij}+2R_{uiuj}(x)\big)\ .
\label{nzero}
}
Notice that the opposite sign of the $b$ coefficient means that scalars 
and spinors respond oppositely to the Ricci curvature. Since the null energy 
condition requires $R_{uu}(x) > 0$, this means the low-frequency phase velocity
is superluminal for spinors in a conformally flat background, but 
subluminal for scalars.

\section{Vacuum Polarization and the Refractive Index}

The propagation of photons at the quantum level 
is determined by the terms in the effective action quadratic in
$A^\mu(x)$. This is the vacuum polarization:
\EQ{
\int
\sqrt{g(x)}\,d^4x\,\sqrt{g(x')}d^4x'\,A^\mu(x)\Pi_{\mu\nu}(x,x')A^\nu(x')
\ .
}
where $g(x)=\det[-g_{\mu\nu}(x)]$. At the one-loop level, the 
on-shell condition for the photon wavefunction is therefore
\EQ{
\nabla_\nu F^\nu{}_\mu=
-4\int \sqrt{g(x')}\,d^4x'\,\Pi^\text{1-loop}_{\mu\nu}(x,x')
A^\nu(x')\ .
\label{erg}
}
To find the refractive index at one-loop order, we substitute the 
tree-level form for the photon wavefunction inside the integral and
take the first term off-shell to give an equation for the unknown set
of functions $\vartheta_{ij}(u;\omega)$ in eq.~\eqref{os}.

Notice, however, that even the effective action computed to all orders in 
the derivative expansion \cite{Shore:2002gw} does not entirely
capture the essential 
physics of high-frequency propagation since, as we have shown in 
ref.\cite{Hollowood:2007ku, Hollowood:2007kt} (see also
\cite{Shore:2002gn,Shore:2003jx}), 
the high-frequency dependence of the refractive index is non-perturbative
in the parameter $\omega^2 R / m^4$. The analysis of vacuum polarization
given here automatically includes this crucial non-perturbative behaviour.

\subsection{Vacuum polarization and the Penrose Limit}

\begin{figure}[ht] 
\centerline{\includegraphics[width=2.5in]{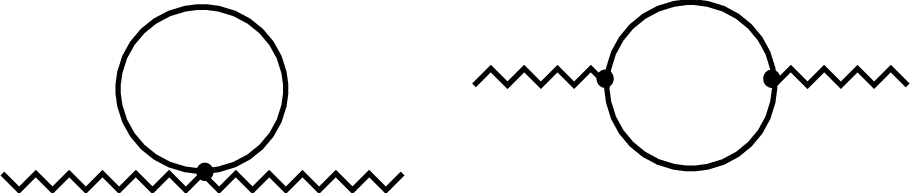}}
\caption{\footnotesize The two Feynman diagrams that contribute to the
vacuum polarization to order $\alpha$.}\label{p5}
\end{figure}

The complete one-loop vacuum polarization 
$\Pi^\text{1-loop}_{\mu\nu}(x,x')$ 
receives contributions from two Feynman diagrams,
as illustrated in Fig.~\ref{p5}.
This gives
\SP{
\Pi^\text{1-loop}_{\mu\nu}(x,x')&=e^2g_{\mu\nu}\delta^{(4)}(x-x')G_F(x,x)\\
&+2e^2\Big[\partial_\mu G_F(x,x')\partial'_\nu G_F(x,x')
- G_F(x,x')\partial_\mu\partial'_\nu G_F(x,x')\Big]\ ,
\label{vacp}
}
where $G_F(x,x')$ is the Feynman propagator of the massive
(scalar) electron. 

The Feynman propagator in a general background spacetime can be written 
in the heat-kernel or ``proper-time'' formalism as
\EQ{
G_F(x,x')=\frac{\sqrt{\det\Delta_{\mu\nu}(x,x')}}{\big(g(x)g(x')\big)^{1/4}}
\int_0^\infty\frac{dT}{(4\pi T)^2}\,ie^{-im^2T+
\tfrac1{2iT}\sigma(x,x')}\Omega(x,x'|T)\ ,
\label{hyu}
}
subject to the usual $m^2\to m^2-i\epsilon$ prescription.
Here, $\sigma(x,x')$ is the geodesic  
interval between the points $x$ and $x'$:
\EQ{
\sigma(x,x')=\tfrac12\int_0^1 d\tau\, g_{\mu\nu}(x)\dot x^\mu\dot x^\nu\
,
\label{intd}
}
where $x^\mu=x^\mu(\tau)$ is the geodesic joining $x=x(0)$ and 
$x'=x(1)$.
The factor $\det\,\Delta_{\mu\nu}(x,x')$ 
is the famous Van Vleck-Morette (VVM) determinant, where the matrix is
\EQ{
\Delta_{\mu\nu}(x,x')= \frac{\partial^2\sigma(x,x')}{\partial x^\mu\partial 
x^{\prime\nu}}\ .
\label{juq}
}
The geometric nature of the VVM matrix and its relation to geodesic deviation
is explored in detail in Section 5.

This expression for the propagator  
has a nice interpretation in the worldline formalism, in which 
the propagator between two points $x$ and $x'$ is determined by a sum
over worldlines $x^\mu(\tau)$ that connect $x=x(0)$ and $x'=x(T)$
weighted by $\exp iS[x]$ where the action is
\EQ{
S[x]=-m^2T+\frac14\int_0^T d\tau\,g_{\mu\nu}(x)\dot x^\mu\dot
x^\nu\ .
\label{wla}
}
\begin{figure}[ht] 
\centerline{\includegraphics[width=2.5in]{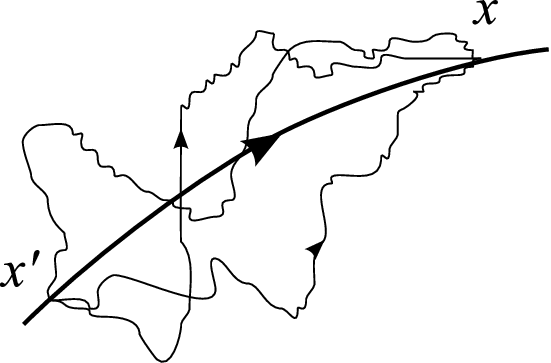}}
\caption{\footnotesize The Feynman propagator $G_F(x,x')$ is expressed as a
  functional integral over paths 
joining $x'$ to $x$. In the limit of weak curvature, $R\ll
  m^2$, the functional integral is dominated by a stationary phase
  solution which is the geodesic joining $x'$ and $x$.}\label{p12}
\end{figure}
Here, $T$ is the worldline length of the loop which is an auxiliary
parameter that must be 
integrated over. The expression \eqref{hyu} corresponds to the
expansion of the resulting functional integral around the 
stationary phase solution, which is simply the classical 
geodesic that joins $x$ and $x'$ as illustrated in Fig.~\ref{p12}. 
In particular, the classical geodesic has an action
$S[x]=\sigma(x,x')/(2T)-m^2T$ giving the exponential terms in
\eqref{hyu}. The VVM determinant comes from integrating
over the fluctuations around the geodesic to Gaussian order while the
term $\Omega(x,x'|T)=1+\sum_{n=1}^\infty a_n(x,x')T^n$ encodes all the
higher non-linear corrections. Notice 
that these terms are effectively an expansion
in $R/m^2$, so the form for the propagator is useful in the limit of
weak curvature compared with the Compton wavelength of the electron.
Of course, this is precisely the limit we are working in here.

\vskip0.3cm
The weak curvature 
limit $R\ll m^2$ leads to a considerable simplification as we now explain.
The terms in the exponent in the second term in \eqref{vacp} are of
the form
\EQ{
\exp\Big[-im^2T-\Big(\frac1{T_1}+\frac1{T_2}\Big)
\frac{\sigma(x,x')}{2i}-i\omega
V'\Big]\ .
\label{zll}
}
For later use, we find it convenient to change variables from $T_1$
and $T_2$ to $T=T_1+T_2$ and
$\xi=T_1/T$, so $0\leq\xi\leq1$. Expressed the other way
\EQ{
T_1=T\xi\ ,\qquad T_2=T(1-\xi)\ .
}
The Jacobian is
\EQ{
\int_0^\infty\frac{dT_1}{T_1^2}\,\frac{dT_2}{T_2^2}=\int_0^\infty
\frac{dT}{T^3}\int_0^1\frac{d\xi}{[\xi(1-\xi)]^2}\ .
}

In the limit $R\ll m^2$ the integral over $x'$ is dominated by a
stationary phase determined by extremizing the exponent \eqref{zll}
with respect to $x'$:
\EQ{
-\frac1{2T\xi(1-\xi)}\partial'_\mu\sigma(x,x')+
\omega\partial'_\mu V'=0\ .
\label{gew}
}
Since $\partial^{\prime\mu}\sigma(x,x')$ is the tangent vector at $x'$
of the geodesic passing through $x'$ and $x$, the stationary phase solution
corresponds to a geodesic with tangent vector $\varpropto
\partial^{\prime \mu}V'$. This means that $x$ and $x'$ must lie on one
of the geodesics of the null congruence. If we choose $x$ to be the
point $(u,0,0,0)$ then  
$x'$ must have Rosen coordinates $(u',0,0,0)$. We call this
distinguished null geodesic $\gamma$. For these points,
it follows that for any metric for which $\partial_V$ is a Killing vector,
\EQ{
\partial_{V'}\sigma(x,x')=u-u'\ ;
}
so the $V'$ component of \eqref{gew} becomes
\EQ{
\frac{u'-u}{2T\xi(1-\xi)}+\omega=0
}
and hence
\EQ{
u'=u-2\omega T\xi(1-\xi)\ .
\label{cdf}
}

\begin{figure}[ht] 
\centerline{\includegraphics[width=2.5in]{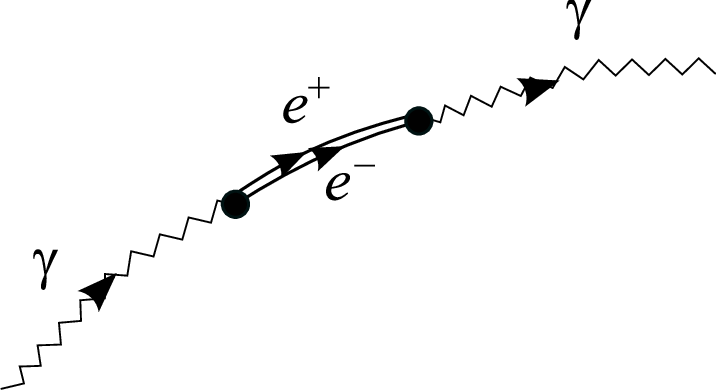}}
\caption{\footnotesize The classical stationary phase solution where
  the photon travelling along the geodesic $\gamma$
decays to an electron-positron pair at $x(u')$ which both follow
  the geodesic $\gamma$ and then  re-combine back into the photon at
  $x(u)$ where $u=u'+2\omega T\xi(1-\xi)$.}\label{p3}
\end{figure}
In the equivalent worldline picture, the stationary phase solution
which dominates in the limit $R\ll m^2$ describes a situation where
the incoming photon decays to an electron positron pair at the point
$u'=u-2\omega T\xi(1-\xi)$ which
propagate along the null geodesic $\gamma$ to the point $u$ and then
combine into the photon again, as shown in Fig.~\ref{p3}. This was a key 
step in the derivation of the refractive index in the worldline formalism
which we presented in ref.\cite{Hollowood:2007kt,Hollowood:2007ku}.

\begin{figure}[ht] 
\centerline{\includegraphics[width=2.5in]{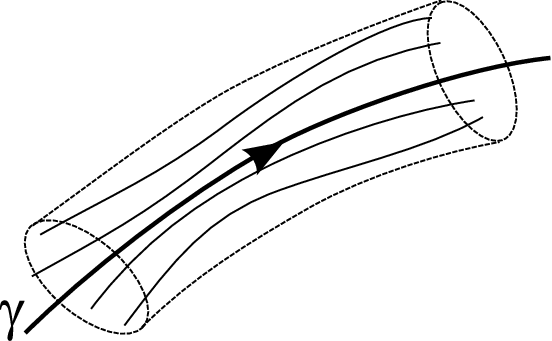}}
\caption{\footnotesize The Penrose limit associated to the null geodesic
  $\gamma$ is the limit of the full metric that captures the tidal
  forces on nearby null geodesics.}\label{p7}
\end{figure}

\vskip0.3cm
In either formalism, the fluctuations around the stationary phase solution 
are governed by the ratio $RT$ which is effectively $R/m^2$. 
In order to set up the expansion systematically 
it is useful to make the following re-scaling of the coordinates
\EQ{
(u, V,Y^a)\longrightarrow(u,T V,\sqrt TY^a)\ .
\label{ps}
}
which implements an overall Weyl scaling while preserving the
stationary phase solution \eqref{cdf}.
After these re-scalings, the geodesic interval \eqref{intd} with
metric \eqref{mrosen} becomes 
\EQ{
\sigma(x,x')=\frac T2\int_0^1d\tau\,\Big[-2\dot u\,\dot  V
+C_{ab}(u,0,0)\dot Y^a\,\dot Y^b\Big]+{\cal O}(T^2)
\label{gact}
}
The leading order piece is precisely the Penrose limit around the null
geodesic $\gamma$ ($V=Y^a=0$). The Penrose limit is the limit of the full
metric in a tubular neighbourhood of a null geodesic, as illustrated in
Fig.~\ref{p7}, defined in such a way that it captures
the tidal forces on the null geodesics that are infinitesimal
deformations of $\gamma$. (This point of view will be described more 
fully in Section~5.) 
It follows that to leading order in the
expansion in $R/m^2$, we can replace the metric by its Penrose limit
around the null geodesic $V=Y^a=0$:
\EQ{
ds^2=-2du\,d V+C_{ab}(u)dY^a\,dY^b\ ,
\label{pel}
}
where $C_{ab}(u)\equiv C_{ab}(u,0,0)$. This defines a plane wave
in Rosen coordinates.

\subsection{Geometry of the plane-wave metric}

The fact that the leading-order contribution to the vacuum polarization
for an arbitrary curved spacetime depends only on its Penrose limit
is a remarkable simplification. As we show below, it allows the derivation of 
a strikingly elegant expression for the full frequency dependence of the 
refractive index, given purely in terms of the VVM matrix.
First, we collect some geometrical properties of the plane wave metric in both 
Rosen and Brinkmann coordinates.

The connection between the Rosen coordinates $(u, V,Y^a)$ and Brinkmann 
coordinates $(u,v,y^i)$ involves a zweibein $E^i{}_a(u)$, which ensures that
the transverse space is flat in Brinkmann coordinates. That is,
\footnote{Note that the $i$ index on $E^i{}_a$ is raised and lowered with
$\delta_{ij}$ while the $a$ index is raised and lowered with
$C_{ab}(u)$ and its inverse. Also note that in Rosen coordinates,
$\sqrt{g(u)} = {\rm det}\, E$.} 
\EQ{
C_{ab}(u)=E^i{}_a(u)\delta_{ij}E^j{}_b{}(u)\ ,
} 
Then, solving the null geodesic equation in the plane wave metric 
\cite{Hollowood:2007ku, Hollowood:2007kt} motivates
the following coordinate transformation:
\SP{
y^i&=E^i{}_a Y^a\ ,\\
v&=V+\tfrac12\frac{dE_{ia}}{du}E^i{}_bY^aY^b \ .
}
The inverse transformations are therefore
\SP{
Y^a &= y^i E_i{}^a\ , \\
V &= v -\tfrac12 \Omega_{ij} y^i y^j \ ,
}
where $\Omega_{ij} = \partial_u E_{ia} E_j{}^a$ plays an important r\^ole in 
the Brinkmann analysis. In particular, the zweibein must be chosen in
such a way that $\Omega_{ij}=\Omega_{ji}$.
In these coordinates, the Penrose limit \eqref{pel} takes the 
familiar plane-wave form
\EQ{
ds^2=-2du\,dv-h_{ij}(u)y^i\,y^j\,du^2+dy^i\,dy^i\ ,
\label{spwmetric}
}
where the quadratic form is
\EQ{
h_{ij}(u) = R_{iuju}= - \frac{d^2E_{ia}(u)}{du^2}E_j{}^a(u)\ ,
}
and we have the useful identity $\Bh = -\partial_u \BOmega -
\BOmega^2$. Here, and in the following boldface symbols are used to
denote $2\times 2$ matrices with Brinkmann transverse $i,j$ indices.

The Brinkmann coordinates are 
more fundamental to the distinguished geodesic
$\gamma$, $Y^a=V=0$, than the Rosen coordinates, since they are the 
geodesic analogues of Riemann normal coordinates known also as 
Fermi normal coordinates \cite{Blau:2006ar}. In addition, 
the Rosen coordinates are not unique since there are always many
inequivalent congruences of which $\gamma$ is a member.  
In the following, we find the Rosen coordinates to be the most 
efficient for performing the calculation while the final result is
naturally expressed in terms of the more fundamental Brinkmann  
coordinates. 

The geodesic interval is particularly simple in Rosen coordinates:
\EQ{
\sigma(x,x')=-(u-u')(V- V')+\tfrac12\Delta_{ab}(u,u')(Y-Y')^a(Y-Y')^b
\label{wee}
}
where
\EQ{
\Delta_{ab}(u,u')=(u-u')\left[\int_{u'}^uC^{-1}(u'')du''\right]^{-1}_{ab}\ .
\label{ddu}
}
$\Delta_{ab}$ are therefore the transverse Rosen components of the 
full VVM matrix. The VVM determinant itself
reduces to a determinant over the two-dimensional transverse space 
\EQ{
\det\,\Delta_{\mu\nu}(x,x')=-\det\,\Delta_{ab}(u,u')\ .
}

To prove this, we first take $\tau=(u''-u')/(u-u')$ in the definition
\eqref{intd} so that
\EQ{
\sigma(x,x')=\frac{u-u'}2\int_{u'}^udu''\,\Big(-2\dot V+C_{ab}\dot
Y^a\dot Y^b\Big)\ .
\label{sigma}
}
The geodesic equation for the $Y^a(u)$ is simply
\EQ{
\frac d{du}C_{ab}(u)\dot Y^b(u)=0\ ,
}
with solution
\EQ{
\dot Y^a(u)=\big[C^{-1}(u)\big]^{ab}\xi_b
}
for constant $\xi_b$. Integrating this, and using the definition
\eqref{ddu}, we have
\EQ{
(Y-Y')^a=\int_{u'}^u du''\,\big[C^{-1}(u'')\big]^{ab}\xi_b=(u-u')
\big[\Delta^{-1}(u,u')\big]^{ab}\xi_b\ .
}
Hence, the geodesic interval is
\SP{
\sigma(x,x')&=-(u-u')(V-V')+\tfrac{u-u'}2(Y-Y')^a\xi_a\\
&=-(u-u')(V-V')+\tfrac12\Delta_{ab}(u,u')(Y-Y')^a(Y-Y')^b
}
as claimed.

In addition, $\Omega(x,x'|T)=1$ in a plane-wave background, which is a
manifestation of the fact that the propagator is WKB exact. This is
entirely consistent with the fact that for the original metric the
non-leading terms in $\Omega(x,x'|T)$ are suppressed in the 
limit $m^2\gg R$. The implication of this is that in a general background
spacetime our analysis is valid in the limits $\omega\gg\sqrt R$ and 
$m^2\gg R$. However, for a plane wave spacetime, the results will actually 
be exact for any $R$, $m$ and $\omega$.

The eikonal approximation \eqref{dds} for the electromagnetic field 
$A_\mu(x)$ is similarly exact for a plane wave spacetime. Moreover, all
the quantities involved have a very simple geometric interpretation
\cite{Hollowood:2007ku,Hollowood:2007kt}.
Specifically, the amplitude is 
\EQ{
{\cal A}(x)=\big(\det\,E_{ia}(u)\big)^{-1/2}
\label{sam}
}
and the non-vanishing components of the polarization vector
$\hat\varepsilon_{(i)\mu}$ are
\EQ{
\hat\varepsilon_{(i)a}(x)=E_{ia}(u)
}
The tree level contribution to the mass shell condition \eqref{erg}
at a point $x=(u,0,0,0)$, for small $\vartheta$, is then simply
\EQ{
\nabla_\nu F_{(i)}{}^\nu{}_a(u)=\omega^2 {\cal A}
(u)\Big[n_{ij}(u;\omega)-\delta_{ij}\Big]E_{ja}(u)\ .
\label{tht}
}

\subsection{Refractive Index}

We now complete the calculation of the vacuum polarization and
refractive index, working from here onwards in a plane wave background.
Returning to the expression \eqref{vacp} for the vacuum polarization,
we find that the first Feynman diagram in Fig.~\ref{p5} gives the 
following contribution to \eqref{erg}: 
\EQ{
\frac{\alpha}{\pi}{\cal A}(u)E_{ia}(u)
\int_0^\infty\frac{dT}{T^2}\,ie^{-im^2T}\ ,
}
By itself, this contribution is divergent but we shall
find that it cancels a divergence in the second term.

The contribution to the on-shell condition \eqref{erg} from the second
Feynman diagram in Fig.~\ref{p5} is then
\EQ{
8e^2\int \sqrt{g(x')}d^4x'\,
\Big[\partial_\mu G_F(x,x')\partial'_\nu G_F(x,x')
- G_F(x,x')\partial_\mu\partial'_\nu G_F(x,x')\Big]\varepsilon^\nu_{(i)}
e^{-i\omega V(x')}\ .
\label{uyt}
}
In Rosen coordinates, we take $x=(u,0,0,0)$ and $x'=(u', V',Y^{\prime a})$.
What remains is to integrate over 
$x^{\prime\mu}$. The integral over $V'$ is trivial and leads to a
delta function constraint
\EQ{
\int dV'\,\exp\Big[\frac{(u'-u)V'}{2iT\xi(1-\xi)}
-i\omega  V'\Big]=4\pi T\xi(1-\xi)\delta\big(u'-u+2\omega T\xi(1-\xi)\big)
\label{yju}
} 
which saturates the $u'$ integral. This simply enforces the condition 
\eqref{cdf}, the stationary phase solution becoming exact for the plane
wave background.

Since $\varepsilon^\nu_{(i)}$ only has non-vanishing components in the
$Y^a$ directions and the integrals over the $Y^{\prime a}$ are Gaussian, 
it follows that \eqref{uyt} is only non-vanishing if the derivatives 
$\partial_\mu$ lie in the directions $\partial_a$.
Using this fact, the $Y^{\prime a}$ integrals are of the form
\EQ{
\int d^2Y'\,\frac{\partial}{\partial Y^{\prime a}}
e^{\tfrac i{4T\xi}Y'\cdot\Delta(u,u')\cdot Y'}\frac{\partial}{\partial
  Y^{\prime b}}
e^{\tfrac i{4T(1-\xi)}Y'\cdot\Delta(u,u')Y'}=
\frac{\pi\xi(1-\xi)}{2}\frac{\Delta_{ab}(u,u')}
{\sqrt{\det\Delta_{ab}(u,u')}}\ .
\label{trans}
}
This is where the advantage of performing the calculation in Rosen coordinates 
is clearest, since these coordinates automatically exhibit the simple form
\eqref{wee} for the transverse sector of the geodesic interval. (The 
corresponding expression in Brinkmann coordinates is given in Section 5.)
Noting that 
\EQ{
\Delta_{ab}(u,u')E_i{}^b(u')=E^j{}_a(u)\Delta_{ij}(u,u')
}
and the fact that 
\SP{
\sqrt{\frac{\det\,\Delta_{ab}(u,u')}{g(u)}}\big(\det\,E_{ia}(u')
\big)^{-1/2}
&=\big(\det\,E_{ia}(u)\big)^{-1/2}\sqrt{\det\,E_i{}^a(u)
\Delta_{ab}(u,u')E_j{}^b(u')}\\ &=
{\cal A}(u)\sqrt{\det\Delta_{ij}(u,u')}\ ,
}
we find the contribution to the mass shell condition \eqref{erg} is
\EQ{
-{\cal A}(u)E^j{}_a(u)
\frac{\alpha}{\pi}\int_0^\infty\frac{dT}{T^2}\int_0^1d\xi
\,ie^{-im^2T}
\Delta_{ij}(u,u')\sqrt{\det\Delta_{ij}(u,u')}
\Big|_{u'=u-2\omega T\xi(1-\xi)}\ .
}

Summing over the two contributions to \eqref{vacp} gives the complete one-loop
term in \eqref{erg} and since the tree-level contribution is
\eqref{tht}, we can extract the matrix of
refractive indices:
\SP{
\Bn(u;\omega)&=\Bone
+\frac{\alpha}{2\pi\omega^2}\int_0^{\infty-i\epsilon}\frac{dT}{T^2}\,
ie^{-im^2T}\\ &\times
\int_0^1d\xi\,\left[\Bone
-\BDelta(u,u')\sqrt{\det\BDelta(u,u')}
\right]_{u'=u-2\omega T\xi(1-\xi)}\ 
\label{refindex}
}
Here, and in the following, $\BDelta$ represents the $2\times2$ matrix
$\Delta_{ij}$ with Brinkmann coordinate indices.

This is our principal result for the refractive index. It is remarkable that
the full frequency dependence of the refractive index/phase velocity for
photons propagating in an arbitrary background spacetime can be expressed
in such a simple and elegant way. The key insight, that the quantum effects 
on photon propagation are determined by the geometry of geodesic fluctuations
around the classical null trajectory and are therefore entirely encoded
in the plane-wave Penrose limit of the original spacetime, explains why
the final result should depend so simply on the VVM matrix only.

Notice that the result for the refractive index at a point $x=x(u)$ 
only depends upon data associated to the
classical null geodesic $x(u-t)$, $0\leq t\leq\infty$, {\it
  i.e.\/}~on the portion in the past relative to $x$. We can write 
\EQ{
\Bn(u;\omega)
=\Bone-\frac{\alpha}{2\pi\omega}\int_0^1
d\xi\,\xi(1-\xi)\BcalF\Big(u;
\frac{m^2}{2\omega\xi(1-\xi)}\Big)\ ,
\label{qya}
}
with
\SP{
\BcalF(u;z)&=\int_0^{\infty-i\epsilon}\frac{dt}{t^2}\,ie^{-izt}\,
\left[
\BDelta\big(u,u-t\big)\sqrt{\det\BDelta\big(u,u-t\big)}-\Bone
\right]\ ,
\label{qybb}
}
where we changed variables from $T$ to 
$t=2\omega\xi(1-\xi)T$. The result we have obtained is strictly valid
for $\omega$ real and positive. Also notice that the definition \eqref{qybb}
has the form of a Fourier Transform of a function which vanishes for $t<0$.

\begin{figure}[ht] 
\centerline{\includegraphics[width=2.5in]{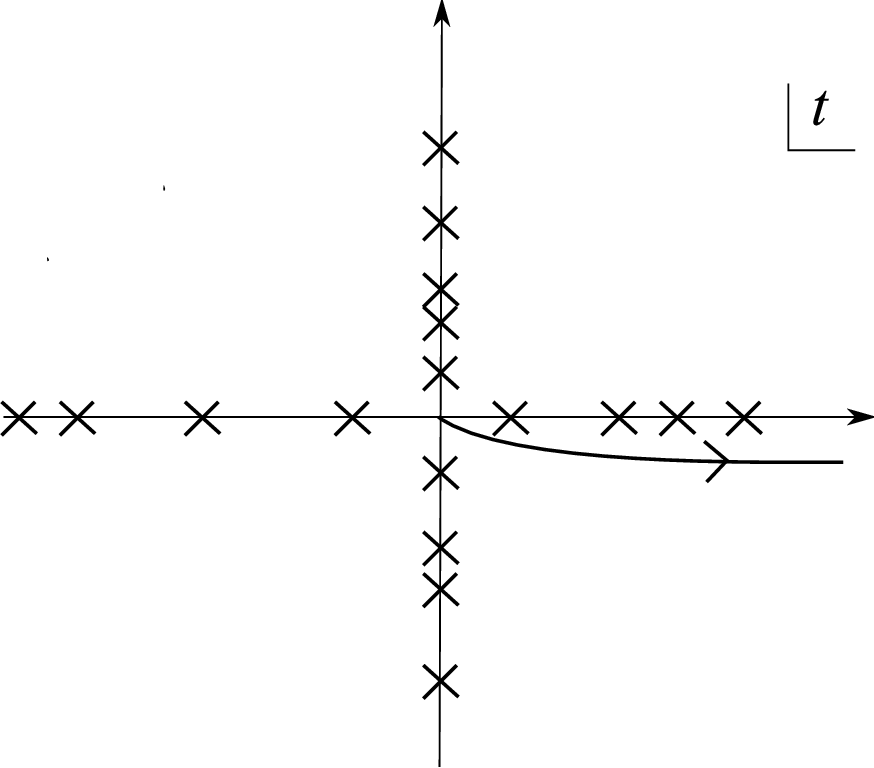}}
\caption{\footnotesize The integration contour in the $t$-plane that
  defines the physical values of the refractive index for real
  positive $\omega$. Crosses represent branch-point or pole 
singularities which generically
  lie on the real axis but in some examples lie also on the imaginary 
axis.}\label{p2}
\end{figure}
We will show in Section~6 that as a consequence of singularities in
$\BDelta$ corresponding to conjugate points on the null congruence,
the integrand has branch-point
singularities on the real $t$ axis and so the $t$ integral must be
defined by some prescription. 
The $i\epsilon$ prescription that we have chosen 
in \eqref{qybb} 
ensures that the refractive index becomes trivial in the flat-space
limit. The $t$ integration contour is illustrated in Fig.~\ref{p2}.

\vskip0.2cm
The low frequency behaviour of the refractive index follows readily
by expanding the VVM matrix in powers of $t$, since the effective
expansion parameter is $\omega^2R/m^4$. Expanding
\EQ{
\BDelta\big(u,u-t\big)\to\Bone+\sum_{n=1}^\infty
\Bs^{(n)}(u)t^{2n}\ ,
}
where $\Bs^{(n)}(u)\varpropto R^n$, we identify the term linear in the 
curvature as 
\EQ{
s_{ij}^{(1)}(u)=-\tfrac16R_{uiuj}(u)\ .
}
Substituting this expansion into \eqref{qya}, we therefore find to leading 
order
\EQ{
n_{ij}(u;\omega)=\delta_{ij}-\frac{\alpha}{360\pi
  m^2}\big(R_{uu}\delta_{ij}-2R_{uiuj}\big)+\frac R{m^2}{\cal O}\Big(
\frac{\omega^2R}{m^4}\Big)\ .
}
in agreement with the result \eqref{nzero} derived from the 
effective action.

\section{Geodesic Deviation and the VVM Determinant}

The Van Vleck-Morette determinant plays a central r\^ole in determining the
refractive index and its analytic structure. This is because the
VVM matrix controls the geometry of geodesic deviation. 
Since this is such an important part of our analysis, in this Section 
we present a detailed account of this geometry, mostly from the viewpoint
of Brinkmann coordinates.

We start with the definition. Fix two points $x'$ and
$x$ in spacetime and consider the following functional integral
\EQ{
\int [dx\,\sqrt{g(x)}]e^{iS[x]}\ ,
}
where the action is
\EQ{
S[x]=\tfrac14\int_0^1 d\tau\,g_{\mu\nu}(x)\dot x^\mu\dot
x^\nu
\label{wlatwo}
}
and the function $x(\tau)$ has boundary conditions $x(0)=x'$ and
$x(1)=x$. The VVM determinant arises from integrating the
fluctuations to Gaussian order around a 
stationary phase solution of the equations of motion. These are
precisely the geodesic equations.  If we denote by $\bar x(\tau)$ 
a (usually unique) 
geodesic that passes through $x'=\bar x(0)$ and $x=\bar x(1)$, the
equations for the fluctuations are
\EQ{
{\cal D}^\mu{}_\nu\delta x^\nu(\tau)=0\ ,
\label{gdone}
}
where ${\cal D}^\mu{}_\nu$ is the second order differential operator
\EQ{
{\cal D}^\mu{}_\nu=\delta^\mu{}_\nu\frac{D^2}{D\tau^2}+
R^\mu{}_{\sigma\nu\lambda}\dot{\bar x}^\sigma\dot{\bar
  x}^\lambda\ .
\label{gdtwo}
}
Here, $D/D\tau$ are absolute derivatives along $\bar x(\tau)$, 
{\it i.e.\/}~$Dx^\mu(\tau)/D\tau=\partial_\tau x^\mu+
\Gamma^\mu(\bar x)_{\nu\sigma}\dot{\bar x}^\nu x^\sigma$. 
The VVM determinant is defined as the functional determinant 
$\det\,{\cal D}^\mu{}_\nu$. It is evaluated directly
in \cite{Papadopoulos:1975ny} (see also \cite{zannias}) 
by discretizing the functional integral and then taking a continuum
limit to yield the finite determinant
\EQ{
\det\,A^\mu{}_\nu(1)\ ,
}
where $A^\mu{}_\nu(\tau)$ is a solution of the Jacobi equation 
\EQ{
{\cal D}^\mu{}_\nu A^\nu{}_\sigma(\tau)=0\ ,
\label{bct}
}
subject to the boundary conditions
\EQ{
A^\mu{}_\nu(0)=0\ ,\qquad
\frac{\partial
  A^\mu{}_\nu(0)}{\partial\tau}=\delta^\mu{}_\nu\ .
\label{bcs}
}

To understand this in more detail and connect to the previous definition
of the VVM determinant, we now specialize to Brinkmann coordinates and 
consider the tidal forces on null geodesics that are infinitesimal 
deformations of the distinguished null geodesic $\gamma$.\footnote{It is 
important to realize that these nearby geodesics do not necessarily lift 
to geodesics of the full metric: this is a global issue of integrability 
that is irrelevant to our discussion.}
These nearby geodesics are described by the 
``Jacobi fields'' in the neighbourhood of $\gamma$ which, given their 
interpretation as Fermi null coordinates, can simply be identified as
the transverse Brinkmann coordinates $y^i(u)$ of null geodesics in the 
plane wave metric \eqref{spwmetric}.
Their evolution is described by the geodesic deviation equation,
specializing \eqref{gdone} and choosing $u$ as the affine parameter:
\EQ{
\frac{d^2 y^i(u)}{du^2}= -R^i{}_{uju}(u)y^j(u)\ .
\label{hyw}
}

The solution of eq.\eqref{hyw} determines the coordinates $y(u)$ in terms
of initial data $y(u')$ and $\dot y(u')$ at a fixed point $u'$, {\it i.e.}
\EQ{
y^i(u) ~=~ B^i{}_j(u,u') y^j(u') ~+~ A^i{}_j(u,u') \dot y^j(u') \ .
\label{aa}
}
It follows immediately that the matrix functions $\BA(u,u')$ and
$\BB(u,u')$, with elements $A_{ij}(u,u')$ and $B_{ij}(u,u')$,
satisfy the geodesic deviation equations
\EQ{
\ddot\BA+\Bh \BA=0 \ ,~~~~~~~~~~~
\ddot\BB+\Bh\BB=0 \ ,
\label{hh}
}
where $\Bh$ has elements $h_{ij}(u)=R_{iuju}$,
with boundary conditions $\BA(u',u') = 0$,
$\partial_u \BA(u,u')|_{u=u'} = \Bone$, $\BB(u',u')=\Bone$ and 
$\partial_u \BB(u,u')|_{u=u'} = 0$.  Using the zweibein, it is easy to see
that these functions satisfy the consistency relation
\EQ{
B_{ij} + A_{ik}\Omega^k{}_j(u') = E_{ia}(u) E_j{}^a(u')\ .
\label{ff}
}

Two special choices of boundary conditions for $y^i(u)$ are of
particular interest: 
\EQ{
{\rm (i)}~~~~~~~~~~~~~~~y^i(u') = 0 ~~~~~\Rightarrow ~~~~~
y^i(u) = A^i{}_j(u,u') \dot y^j(u') \ ,
\label{cc}
}
with, as always, $A_{ij} =0$ and $\partial_u A_{ij} = \d_{ij}$ at $u=u'$.
This describes a ``spray'' of geodesics \cite{Visser:1992pz} passing through a 
point $y(u') = 0$ and determines the function $A_{ij}$ which, 
as we show below, 
is related very simply to the inverse of the VVM matrix $\Delta_{ij}$. In
addition, as we prove below, $\BA(u,u')$ has the anti-symmetric property
\EQ{
\BA(u,u')=-\BA(u',u)^\top\ .
\label{kq}
}

\EQ{
{\rm (ii)}~~~~~~~~~~~~~~~\dot y^i(u') = 0 ~~~~~\Rightarrow ~~~~~
y^i(u) = B^i{}_j(u,u') y^j(u') \ ,
}
with $B_{ij} =\d_{ij}$ and $\partial_u B_{ij} = 0$ at $u=u'$.
This is the choice \cite{HawkingEllis} appropriate to a geodesic congruence
with neighbouring geodesics parallel at $u'$.

\vskip0.2cm
The geodesic deviation functions $\BA$ and $\BB$ determine the geodesic
interval for the plane wave in Brinkmann coordinates. Analogous to the Rosen
expression \eqref{sigma}, we have
\SP{
\sigma(x,x') &= -\frac{u-u'}{2}\int_{u'}^u du''\,\Big(2\dot v
+ h_{ij}(u'') y^i y^j - \dot y^i \dot y_i \Big)\\
&= -(u-u')\Bigl( (v-v') - \tfrac12 \bigl[\dot y^i y_i\bigr]_{u'}^u \Bigr)\ ,
}
using the geodesic equation $\ddot y^i + h^i{}_j y^j = 0$.
Substituting \eqref{aa} now gives
\SP{
&\sigma(x,x') = -(u-u') \Bigl( (v-v') ~+~
y^i(u)A^{-1}_{ji}(u,u')y^j(u') \\
&~~~~+~\tfrac12 y^i(u) A^{-1}_{ik}(u',u)B^k{}_j(u',u) y^j(u) 
~-~\tfrac12 y^i(u') A^{-1}_{ik}(u,u')B^k{}_j(u,u') y^j(u')~\Bigr)\ .
\label{dd}
}
The transverse Brinkmann components $\Delta_{ij}$ of the VVM matrix,
defined by 
\EQ{
\Delta_{ij}(u,u') =  \frac{\partial^2 \sigma(x,x')} 
{\partial y^i(u) \partial y^j(u')} 
}
and therefore
\EQ{
\BDelta(u,u') = (u-u') \big(\BA^{-1}(u,u')\big)^\top\ .
\label{ee}
}

Yet another interpretation 
\cite{Visser:1992pz} of the VVM determinant is as
the Jacobian for the change of variables between specifying a geodesic
by giving two points---$x(u')$ and $x(u)$---through which it passes
and giving one point and the tangent vector at that point: $x(u')$ and
$\dot x(u')$. Since we can think of
$k^\nu(x,x')=\partial\sigma(x,x')/\partial x^{\prime\nu}$ as the
tangent vector at $x'$ of the geodesic that goes through the two
points $x'$ and $x$, normalized so that the affine parameter between
$x'$ and $x$ goes from $0$ to $1$, we see from \eqref{juq} that
\EQ{
\Delta_\mu{}^\nu(x,x')=\partial_\mu k^\nu(x,x')\ .
} 
With this normalization,
\EQ{
k^i(u,u')=(u-u')\dot y^i(u')
}
and so from \eqref{cc} we have the Jacobian matrix
\EQ{
\frac{\partial k_j(u,u')}{\partial y^i(u)}
=(u-u')A^{-1}_{ji}(u,u')\ .
\label{ll}
}

The equivalence of the Rosen and Brinkmann expressions \eqref{wee} and 
\eqref{dd} for the geodesic interval is readily established once the
following identity is proved:
\EQ{
A_{ij}(u,u') ~=~ E_{ia}(u') \int_{u'}^udu''\,
\big[C^{-1}(u'')\big]^{ab}\,
E_{jb}(u) \ .
\label{gg}
}
The proof is as follows. Notice that the zweibein $E_{ia}(u)$ is a
particular solution of the geodesic equation \eqref{hh}. 
$\BA(u,u')$ also solves this equation and so it
follows that
\EQ{
\BE^\top\dot\BA-\dot\BE^\top\BA=\BK\ ,
}
where $\BK$ is a constant matrix. Using the fact that
$\BOmega=\dot\BE\BE^{-1}$ is a symmetric matrix allows us to write
\EQ{
\partial_u\big(\BE^{-1}\BA\big)=\BE^{-1}\big(\BE^{-1}\big)^\top\BK
=\BC^{-1}\BK\ ,
}
where $\BC(u)$ is the non-trivial part of the 
metric in Rosen coordinates
\eqref{pel}. Integrating this equation and imposing the boundary
conditions $\BA(u',u')=0$ and $\partial_u\BA(u,u')|_{u=u'}=1$, gives
\EQ{
\BA(u,u')=\BE(u)\int_{u'}^u du''\,\BC^{-1}(u'') \BE(u')^\top\ .
}
which in components is \eqref{gg}. Notice that the symmetry \eqref{kq}
is manifest. A similar construction with the alternative boundary 
conditions determines $\BB(u,u')$ in the form \eqref{ff}.\footnote{
Also notice that if we were to evaluate the vacuum 
polarization directly using the Brinkmann expression for $\sigma(x,x')$
rather than the simpler Rosen form, \eqref{ff} is essential in
simplifying the 
transverse integrals and ensuring that the elegant Rosen result 
\eqref{trans} is reproduced.}

\vskip0.2cm
Finally, we relate these results to the optical scalars in the Raychoudhuri 
equations which describe the geodesic flow. It is convenient to start from
an alternative, but entirely equivalent, description of geodesic deviation.
In this approach, the evolution of the Jacobi fields $y^i(u)$ is determined 
by requiring that their Lie derivative ${\cal L}_k y^i$ vanishes along the
geodesic with tangent vector $k^\mu$, {\it i.e.\/}
\EQ{
{\cal L}_k y^i  \equiv k\cdot\nabla y^i - (\nabla_j k^i) y^j = 0 \ .
}
This implies the parallel transport equation 
\EQ{
\partial_u y^i = \Omega^i{}_j y^j
}
where $\Omega_{ij} = \nabla_j k_i$. Notice that since the geodesic tangent 
vector in Brinkmann coordinates is 
$k^\mu = (1, (\frac{1}{2}\dot\Omega + \Omega^2)_{ij}y^i y^j, \Omega_{ij}y^j)$,
this is consistent with the original definition $\Omega_{ij} = \dot E_{ia}
E_j{}^a$. It then follows from \eqref{aa} that
\EQ{
\partial_u \log\bigl(B_{ij}(u,u') + A_{ik}(u,u')\Omega^k{}_j(u')\bigr) ~=~
\Omega_{ij}(u)\ ,
\label{kkk}
}
which is clearly consistent with \eqref{ff}.

The matrix $\Omega_{ij}$ is the fundamental object from which the optical
scalars are defined. We have\footnote{Here, we follow the conventions
of Wald \cite{Wald:1984rg}. There are therefore some factors of 2 different
from the Chandrasekhar \cite{Chandra} conventions used in 
refs.\cite{Hollowood:2007ku,Hollowood:2007kt}.}
\EQ{
\Omega_{ij}(u) = \tfrac12\hat\theta(u) \d_{ij} + \hat\sigma_{ij}(u)
+ \hat\omega_{ij}(u) \ ,
} 
defining the {\it expansion} $\hat\theta$, the {\it shear} $\hat\sigma_{ij}$
and the {\it twist} $\hat\omega_{ij}$. The corresponding scalars are 
$\hat\sigma^2 = \Omega_{(ij)}\Omega^{ij} - \tfrac12\hat\theta^2$  
and $\hat\omega^2 = \Omega_{[i,j]}\Omega^{ij}$. 
The twist vanishes in all cases considered here, so $\Omega_{ij}$ is symmetric.
Eq.\eqref{kkk} therefore implies:
\EQ{
B_{ij}(u,u') + A_{ik}(u,u') \Omega^k{}_j(u') ~=~ \exp \int_{u'}^u du''
\bigl(\tfrac12\hat\theta \d_{ij} + \hat\sigma_{ij}\bigr) \ .
}

Note that the optical scalars depend on the choice of boundary conditions
imposed on $y^i(u)$. A particularly relevant choice is the ``geodesic spray''
condition considered above. In this case, \eqref{kkk} simplifies to 
\EQ{
\partial_u \log\BA ~=~ \BOmega(u) \ .
}
Taking the trace gives the important identity
\EQ{
\partial_u \log \det \BA(u,u') = \hat\theta (u,u') \ ,
}
where we display the $u'$ dependence on the r.h.s.~explicitly as a reminder
of the choice of boundary condition, just as in the notation $k^i(u,u')$
in \eqref{ll} for the tangent vector.

\vskip0.3cm
\begin{figure}[ht] 
\centerline{\includegraphics[width=2.5in]{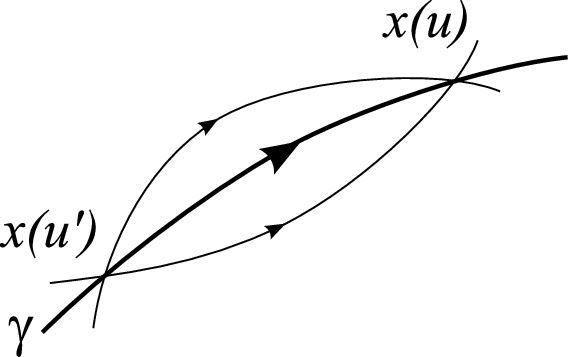}}
\caption{\footnotesize Conjugate points $x(u)$ and $x(u')$ on the
  geodesic $\gamma$ through which infinitesimal deformations of
  $\gamma$ (or finite deformations in the Penrose limit) also pass.}
\label{p11}
\end{figure}

In general, if there are two points $x(u)$ and $x(u')$ on a geodesic 
$\gamma$ for which there exists a family of geodesics infinitesimally 
close to $\gamma$ 
which also pass through $x(u)$ and $x(u')$, then these are said to be 
{\it conjugate points}. As we now show, conjugate points play a crucial 
r\^ole in determining the analyticity properties of the refractive index. For the plane wave, conjugate points correspond to solutions of the geodesic equation
with $y^i(u) = y^i(u') = 0$. It follows from the discussion above that
this implies $\det \BA(u,u') = 0$. In turn, this implies that at these points, the VVM determinant $\det \BDelta(u,u')$ has a singularity. This establishes a direct link between the analyticity structure of the refractive index \eqref{refindex} and the geometry of conjugate points. Moreover, this geometry is entirely encoded in the geodesic deviation matrix $\BA(u,u')$ or equivalently the VVM matrix $\BDelta(u,u')$.

\begin{figure}[ht] 
\centerline{\includegraphics[width=2.5in]{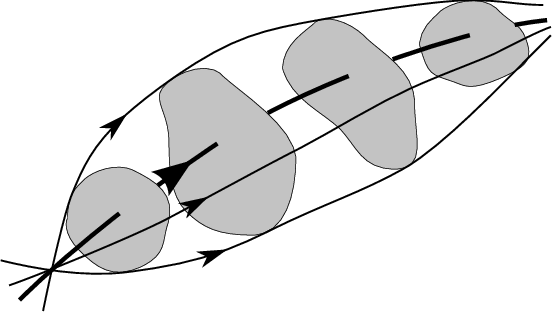}}
\caption{\footnotesize The expansion $\hat\theta(u,u')$ 
describes the rate-of-change of the proper area defined by the spray
  of null geodesics that pass through $x(u')$. 
The null energy condition implies that the expansion monotonically
  decreases, except at conjugate points.}\label{p8}
\end{figure}

A particularly important observation is that the existence of conjugate points is generic (see {\it e.g.} \cite{Wald:1984rg}). This follows from the Raychoudhuri equations:
\SP{
\partial_u \hat\theta &= -\tfrac12\hat\theta^2 - 
\hat\sigma_{ij}\hat\sigma^{ij}
- R_{uu} \\
\partial_u \hat\sigma_{ij} &= -\hat\theta \hat\sigma_{ij} - C_{uiuj} \ .
\label{mm}
}
As a consequence of the null energy condition, $R_{uu}\geq0$, \eqref{mm}
implies the inequality
\EQ{
\partial_u\hat\theta(u,u')+\tfrac12\hat\theta(u,u')^2\leq0\ .
\label{dsq}
}
The significance of this is that $\partial_u\hat\theta(u,u')\leq0$ so that
$\hat\theta(u,u')$ generally decreases monotonically with $u$. 
(Of course, this is violated at the singularities where 
$\hat\theta(u,u')$ jumps from $-\infty$ to $\infty$.) If at some point 
$u=\tilde u$, $\hat\theta(\tilde u,u')$ is negative, say
$-|\lambda|$, then
inevitably $\hat\theta(u,u')\to-\infty$ at some finite $u\leq \tilde
u+2/|\lambda|$. The proof is simple. In order to attempt to avoid the
singularity $|\partial_u\hat\theta(u,u')|$ should be as small as
possible. In other words, we should saturate the inequality \eqref{dsq},
with the solution
\EQ{
\hat\theta(u,u')=\frac2{u-\tilde u-2/|\lambda|}\ .
}
Hence, there must be a conjugate point at some $x(u)$ with $u\leq
\tilde u+2/|\lambda|$. At the conjugate point, $\hat\theta(u,u')$ jumps
discontinuously 
from $-\infty$ to $\infty$ and then begins its descent again.
Notice that as $u\to\infty$ the expansion must go asymptotically to zero.
\begin{figure}[ht] 
\centerline{\includegraphics[width=2.5in]{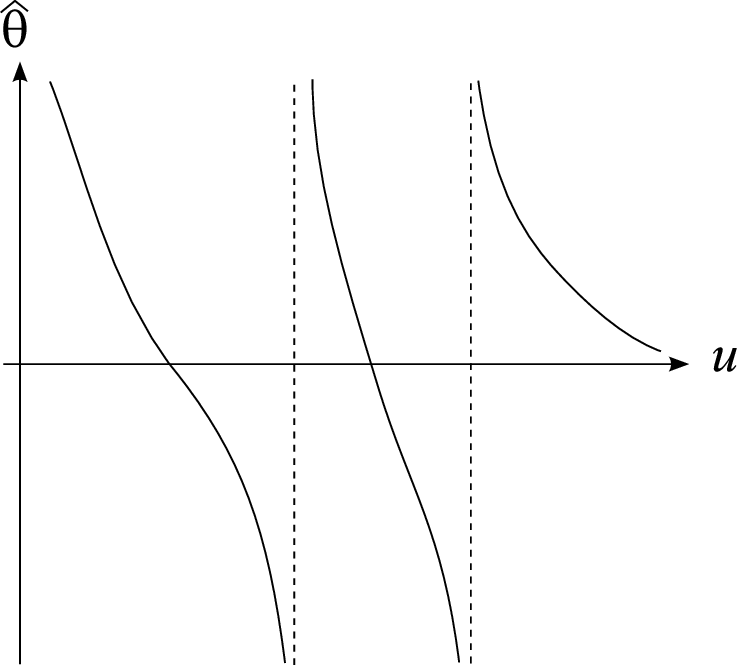}}
\caption{\footnotesize The generic behaviour of the {\it expansion\/}
  $\hat\theta(u,u')$ along a
  null geodesic. The singularities occur when $x(u)$ and $x(u')$ are
  conjugate points.}\label{p10}
\end{figure}

Finally, notice that by diagonalizing the shear tensor and combining 
the two Raychoudhuri equations \eqref{mm}, we can characterize
the null congruences by whether
the geodesics focus in both transverse directions 
(specified by the eigenvectors of $\sigma_{ij}$) or have  
one direction focusing and one defocusing. (These were labelled
as Type I and Type II respectively in 
refs.\cite{Hollowood:2007ku,Hollowood:2007kt}.)
As shown above, the null energy condition prohibits the existence
of a third case with defocusing/defocusing. The focusing 
directions give rise to conjugate points and corresponding 
singularities of the VVM determinant $\BDelta(u,u')$ on the real 
axis; defocusing directions, on the other hand, are associated 
with singularities on the imaginary axis, as illustrated in the 
example of general symmetric plane waves in Section 7.2.

\section{Analyticity and Causality}

As noted at the end of Section 4, the singularities in the VVM 
determinant induced by the existence of conjugate points gives rise
to a novel analytic structure for the refractive index in the
complex $\omega$ plane. This is a generic effect which will also
affect more general scattering amplitudes. As we shall see, it means 
that in curved spacetime some of the conventional axioms and 
assumptions of $S$-matrix theory and dispersion relations 
need to be re-evaluated, with far-reaching physical implications.

\subsection{Analytic structure of the refractive index}

Returning now to the expression \eqref{qya} for the 
refractive index in terms of the VVM matrix, 
\EQ{
\Bn(u;\omega)
=\Bone-\frac{\alpha}{2\pi\omega}\int_0^1
d\xi\,\xi(1-\xi)\BcalF\Big(
\frac{m^2}{2\omega\xi(1-\xi)}\Big)\ ,
\label{ana}
}
it is clear that the $t$-integral defining $\BcalF(u;z)$ in \eqref{qybb}
has branch-point singularities on the real $t$-axis 
whenever $x(u)$ and $x(u-t)$ are conjugate points. After integration
over $t$, these singularities give rise to cuts in $\BcalF(u;z)$
from $0$ to $\infty$ in the complex $z$-plane. The refractive index
$\Bn(u;\omega)$ is therefore a multi-valued function of $\omega$
with branch points at $0$ and $\infty$.

The presence of branch-point singularities on the real $t$-axis means
we have to give a prescription for the contour of the $t$-integration
in \eqref{qybb}. For $z$ real and positive, as in Section 4, we define:
\EQ{
\BcalF_+(u;z)\Big|_{z\in{\bf R}>0}=
\int_0^{\infty - i\epsilon}\frac{dt}{t^2}\,ie^{-izt}\left[\BDelta\big(u,u-t\big)\sqrt{\det\BDelta\big(u,u-t\big)}-\Bone\right]\ ,
\label{anb}
}
with the contour as illustrated in Fig.~\ref{p2}. With this choice,
the $t$-integral can be performed by rotating to the negative imaginary
$t$-axis where the integral is convergent due to the damping of 
$e^{-izt}$ as $t\rightarrow -i\infty$ for Re $z >0$.
Moreover, since the integral is over Re $t >0$, it receives support only
from that part of the null geodesic to the {\it past} of $x(u)$, i.e.~from
$\BDelta(u,u-t)$ with Re $t >0$. Intuitively this is what one would
expect for in a causal theory.

Similarly, for $z$ real and negative we should define
\EQ{
\BcalF_-(u;z)\Big|_{z\in{\bf R}<0}=
\int_0^{-\infty + i\epsilon}\frac{dt}{t^2}\,ie^{-izt}\left[\BDelta\big(u,u-t\big)\sqrt{\det\BDelta\big(u,u-t\big)}-\Bone\right]\ .
\label{anc}
}
This contour avoids the singularities on the negative $t$-axis,
while rotation of the contour towards the positive imaginary $t$-axis
leads to a convergent integral for Re $z < 0$. 
This time, the $t$-integral has support only from the section of the null 
geodesic in the {\it future} of $x(u)$. Again, this is as required for
causality with $\text{Re}\,\omega < 0$ and is consistent with the usual 
flat-space limit. 

The next step is to specify the physical sheet for the multi-valued 
function $\BcalF(u;z)$ defining the refractive index. First, we choose
to run the branch cuts in the complex $z$-plane from 0 to $\pm \infty$
just above (below) the positive (negative) real axis 
respectively.\footnote{This applies also to the cuts arising from any
branch-point singularities occurring off the real $t$-axis, for example
the singularities on the imaginary $t$-axis in the general symmetric plane
wave example discussed in Section 7.}
The physical $\BcalF(u;z)$ is defined as the analytic continuation 
of $\BcalF_+(u;z)$ from real, positive $z$ into the lower-half plane
and of $\BcalF_-(u;z)$ from real, negative $z$ into the upper-half plane.
That is,
\SP{
\BcalF(u;z) ~~ = ~~\begin{cases}
\BcalF_+(u;z) ~~~~~~~~~~~~~~~~-\pi < {\rm arg~}z \le 0 \ , \\
\BcalF_-(u;z) ~~~~~~~~~~~~~~~~~~~~0 < {\rm arg~}z \le \pi \ . \end{cases}
\label{and}
}

\begin{figure}[ht] 
\centerline{\includegraphics[width=2.5in]{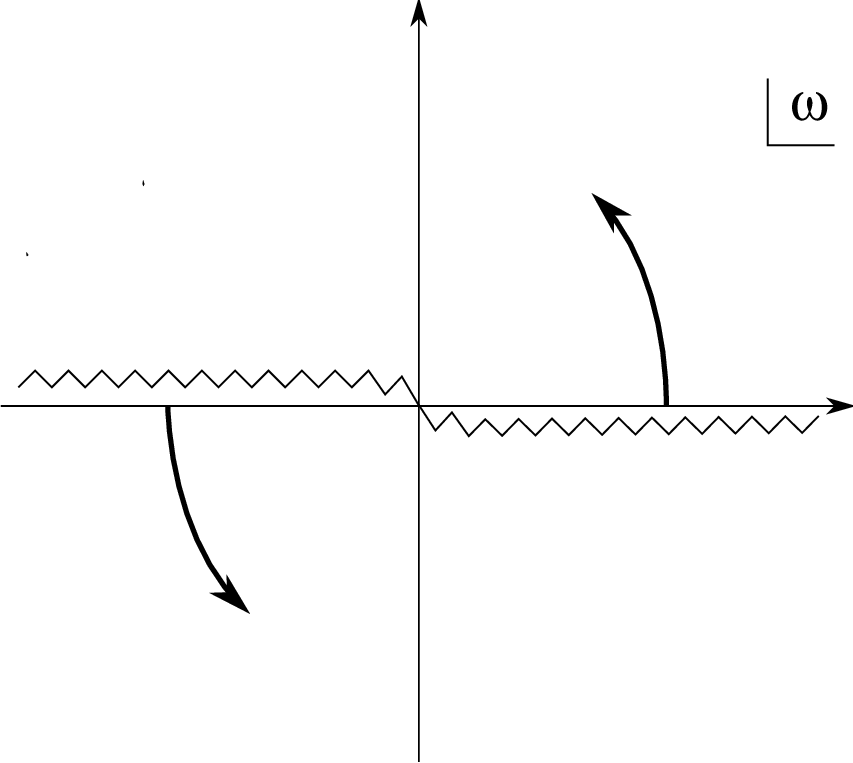}}
\caption{\footnotesize The physical sheet for $\Bn(u;\omega)$
  defined by analytic
  continuation from the real positive axis into the upper-half plane
  and from the real negative axis into the lower-half plane. Since $z$
  is inversely related to $\omega$ the 
  upper-half of the $\omega$ plane maps to the lower-half of the $z$
  plane, and vice-versa.}\label{p6}
\end{figure} 
This implies the corresponding analytic structure for the refractive 
index $\Bn(u;\omega)$ itself in the complex $\omega$-plane,
illustrated in Fig.~\ref{p6}. Since $z$ is essentially the inverse of 
$\omega$, the upper-half plane in $z$ maps into the lower-half plane
in $\omega$ and {\it vice-versa\/}. The physical refractive index is 
therefore given by the analytic continuation of
$\Bn_+(u;\omega)$---defined 
using $\BcalF_+(u;z)$---into the upper-half plane
and of $\Bn_-(u;\omega)$ into the lower-half plane

There may also be further singularities in $\BcalF_+(u;z)$ or 
$\BcalF_-(u;z)$ individually ({\it e.g.}~we will find examples in the
next section where $\BcalF_+(u;z)$ has poles on ${z\in{\bf R}<0}$)
but these lie off the physical sheet defining $\BcalF(u;z)$ itself.

\vskip0.2cm
Across the cuts, there will be a discontinuity in $\BcalF(u;z)$ or
$\Bn(u;\omega)$ and we define Disc $\BcalF(u;z)$ and Disc $\Bn(u;\omega)$
as the discontinuities across the appropriate cuts taken in the 
anti-clockwise sense. These discontinuities play an important r\^ole in 
dispersion relations.

\begin{figure}[ht] 
\centerline{\includegraphics[width=5in]{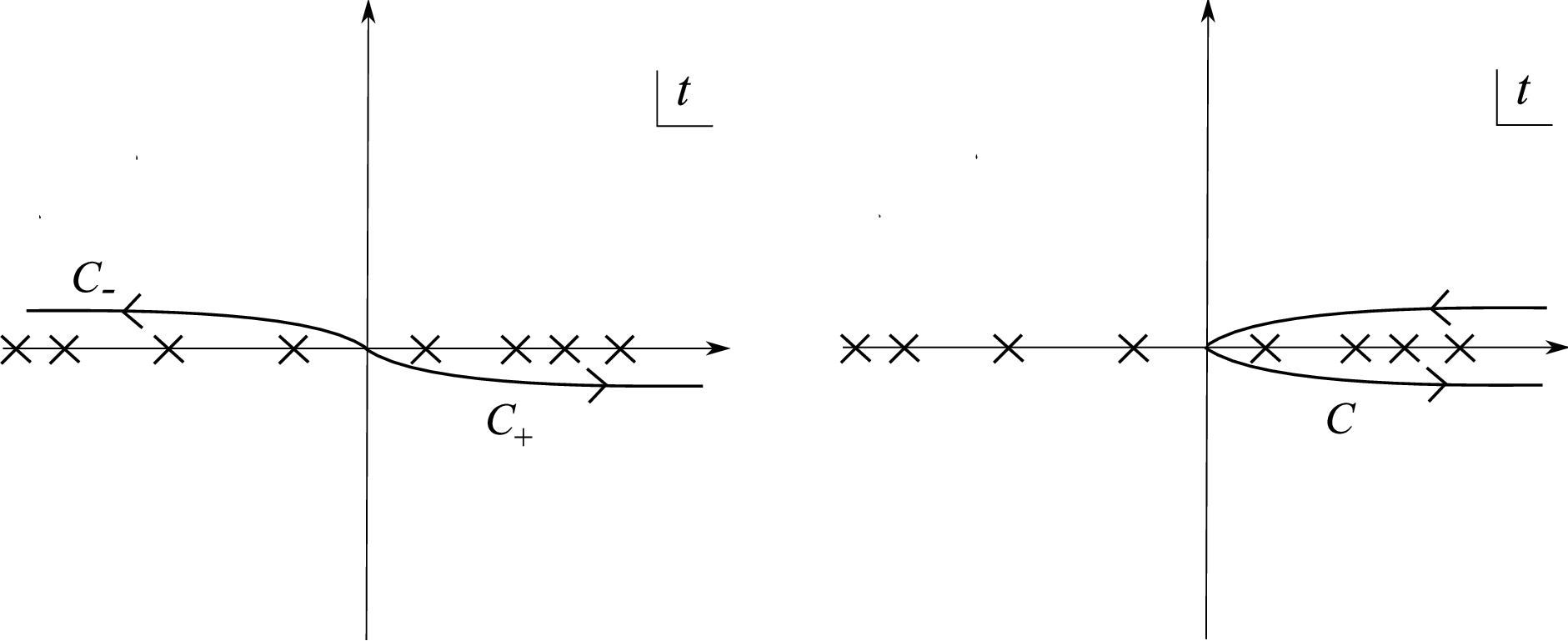}}
\caption{\footnotesize Wrapping the contours ${\cal C_-}$ and ${\cal C_+}$
defining $\BcalF_-(u;\omega)$ and $\BcalF_+(u;\omega)$ around the branch
point singularities on the positive real $t$-axis. The integral with the
 resulting contour ${\cal C}$ gives the discontinuity of $\BcalF(u;\omega)$
across the cut along the positive real $\omega$-axis.}
\label{p21}
\end{figure}

In the simplest case, where there are no singularities in the complex 
$t$-plane apart from those on the real axis (as realized in the 
conformally flat symmetric plane wave example discussed in Section 7.1),
we can evaluate Disc $\BcalF(u;z)$ across the cut along ${z\in{\bf R}>0}$
by rotating the contour defining $\BcalF_-(u;z)$ to wrap around the 
positive $t$-axis as shown in Fig.~\ref{p21}.
That is,
\SP{
{\rm Disc}~ \BcalF(u;z)\big|_{z\in{\bf R}>0}~~~&=~~~
\BcalF_-(u;z) - \BcalF_+(u;z) \\
&=~~~\int_0^{-\infty + i\epsilon} {dt\over t^2}\, i e^{-izt} 
\bigl[\ldots\bigr]
~~-~~\int_0^{\infty - i\epsilon} {dt\over t^2}\, i e^{-izt} 
\bigl[\ldots\bigr] \\
&=\int_{\cal C} {dt\over t^2}i e^{-izt}
\left[\Bone
-\BDelta\big(u,u-t\big)\sqrt{\det\BDelta\big(u,u-t\big)}
\right]\ .
\label{ane}
}
Indeed, for the conformally symmetric plane wave background, the singularities 
on the real $t$-axis are actually poles, so 
Disc $\BcalF(u;z)$ can be evaluated from the contour ${\cal C}$ simply 
as the sum of the residues. This example is worked out
explicitly in Section 7.1.

\vskip0.2cm
An important special case arises when the background is translation invariant
 with respect to the coordinate $u$ along the null geodesic.
Since the VVM matrix is symmetric in its two arguments, we then have
\EQ{
\BDelta(u,u-t) ~~=~~ \BDelta(u-t,u) ~~=~~ \BDelta (u,u+t) \ ,
\label{anf}
}
so the integrand $\bigl[\ldots \bigr]$ in $\BcalF(z)$ (which is of course then 
independent of $u$) is an
{\it even} function of $t$.  In turn, this implies
\EQ{
\BcalF_-(z) ~~=~~ - \BcalF_+(-z) ~~~~~~~~~~~
({\rm translation~invariance})
\label{ang}
}
which is a result of special significance in the 
analysis of dispersion relations.

\subsection{Kramers-Kronig dispersion relation}

The Kramers-Kronig dispersion relation is an identity satisfied by the
refractive index, or vacuum polarization, in QED.\footnote{Contrary  
to some recent claims in the literature \cite{Dubovsky:2007ac}, the
Kramers-Kronig relation is equally valid in relativistic quantum field
theory as it is in non-relativistic settings; for example the proof in
QFT is presented in Weinberg's textbook \cite{Weinberg}.}
Its derivation depends critically on the analyticity properties of the 
refractive index and shows in a simple context the sort of changes to 
conventional $S$-matrix relations and dispersion relations which will
occur due to the novel analytic structure of amplitudes in curved 
spacetime.

\begin{figure}[ht] 
\centerline{\includegraphics[width=5in]{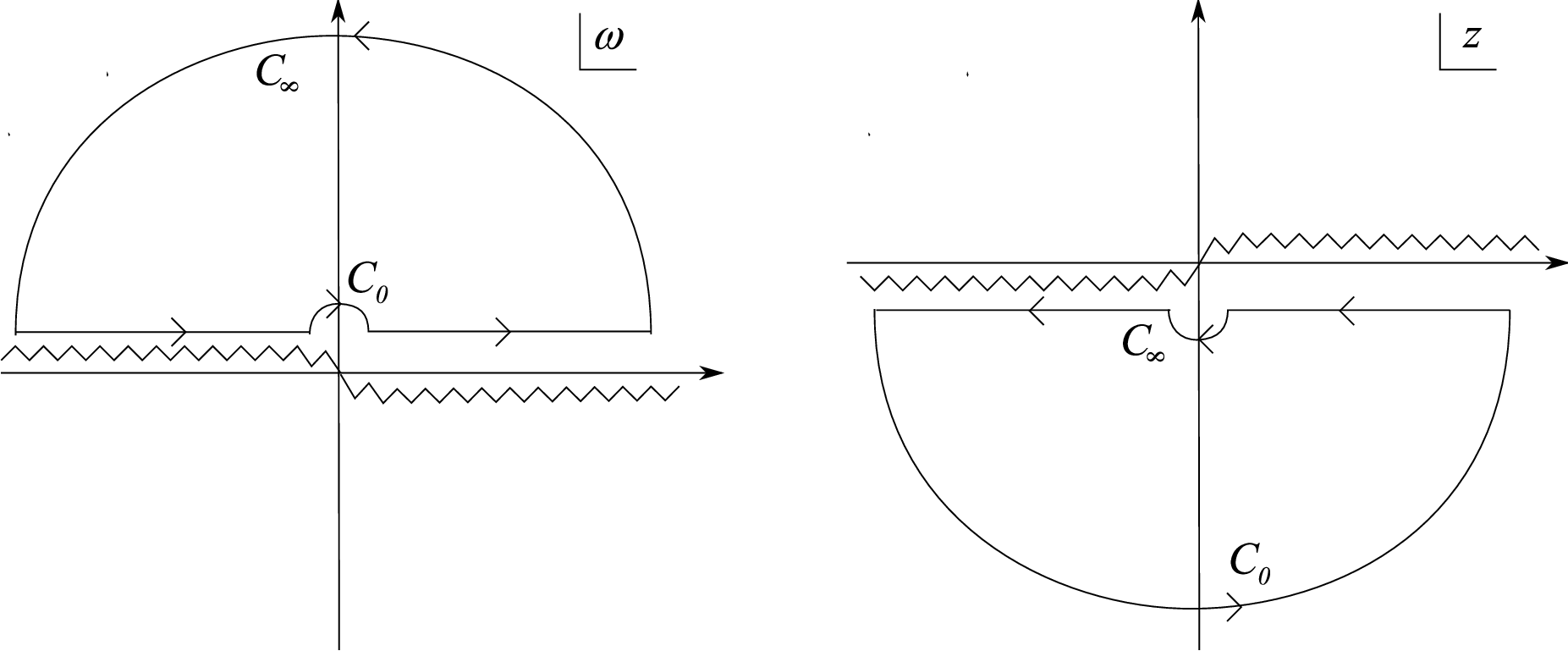}}
\caption{\footnotesize The contour in the complex $\omega$-plane used 
in the derivation of the Kramers-Kronig relation for $\Bn(u;\omega)$. 
The second figure shows the equivalent contour in the complex $z$-plane 
relevant for $\BcalF(u;z)$.}
\label{newfig2}
\end{figure} 
To derive the Kramers-Kronig relation, we integrate $\Bn(u;\omega)/\omega$
around the contour shown in Fig.~\ref{newfig2}. As explained in the 
following section, causality imposes two fundamental properties 
of the refractive index:~(i)~$\Bn(u;\omega)$ is analytic in the upper-half
$\omega$-plane, and ~(ii)~$\Bn(u;\omega)$ is bounded at
infinity.\footnote{Notice that this is weaker than the condition that
  $\Bn(u;\omega)\to\Bone$ as $|\omega|\to\infty$.} 
Assuming these properties, we have
\EQ{
\int_{{\cal C}_\infty}\frac{d\omega}\omega\,\Bn(u;\omega) ~+~
\int_{{\cal C}_0}\frac{d\omega}\omega\,\Bn(u;\omega) ~+~
{\cal P}\int_{-\infty}^\infty \frac{d\omega}\omega\,
\Bn(u;\omega) ~~=~~0 \ .
\label{anh}
}
which implies
\EQ{
\Bn(u;0)-\Bn(u;\infty) ~~=~~
{1\over i\pi}~ {\cal P}\int_{-\infty}^\infty \frac{d\omega}\omega\,
\Bn(u;\omega)\ .
\label{ani}
}
Provided the causality properties are satisfied, the Kramers-Kronig 
relation in the form \eqref{ani} is always valid. Note that the principal 
part integral is over a contour lying just above the real axis, so can
be written as
\SP{
{\cal P}\int_{-\infty}^\infty \frac{d\omega}\omega\, \Bn(\omega) ~~&=~~
\int_{-\infty}^0 \frac{d\omega}\omega\,\Bn(u;\omega + i\epsilon) ~+~
\int_0^\infty\frac{d\omega}\omega\,\Bn(u;\omega + i\epsilon) \\
&=~~\int_0^\infty\frac{d\omega}\omega\,\big(\Bn(u;\omega + i\epsilon)
- \Bn(u;-\omega + i\epsilon)\big) \ . 
\label{anj}
}

Now, in the conventional flat-spacetime derivation \cite{Weinberg},
translation invariance implies that $\Bn(\omega)$ is an {\it even} 
function of $\omega$. This implies that $\Bn(-\omega + i\epsilon)
= \Bn(\omega - i\epsilon)$, so the r.h.s.~of \eqref{anj} becomes
the discontinuity of $\Bn(\omega)$ on the positive real $\omega$-axis:
\EQ{
\Bn(0)-\Bn(\infty) ~~=~~
{1\over i\pi} \int_0^\infty \frac{d\omega}\omega\,
\text{Disc}\, \Bn(\omega)~~~~~~~~~~~~~~~
({\rm translation~invariance}) \ .
\label{ank}
}
Finally, in flat-spacetime QED, the refractive index satisfies the 
property of {\it real analyticity}, $\Bn(\omega^*) = \Bn(\omega)^*$.
This is a special case of the basic $S$-matrix property of
hermitian analyticity \cite{Olive} which is satisfied by more general
scattering amplitudes. With this assumption, we can replace the 
discontinuity in \eqref{ank} by the imaginary part of the refractive
index, since for real $\omega$ it implies $\Bn(\omega - i\epsilon)
= \Bn(\omega + i\epsilon)^*$, leaving
\EQ{
\Bn(0)-\Bn(\infty) ~~=~~
{2\over \pi} \int_0^\infty \frac{d\omega}\omega\,
\text{Im}\,\Bn(\omega)~~~~~~~~~~~~~~~
({\rm real~analyticity}) \ .
\label{anl}
}
This is the standard form of the Kramers-Kronig relation. Since the
optical theorem relates the imaginary part of forward scattering amplitudes
 to the total cross section, the r.h.s.~of \eqref{anl} is positive under
conventional QFT conditions. This would imply $n(0) > n(\infty)$, consistent
 with a subluminal low-frequency phase velocity, which is the 
usual dispersive situation.\footnote{For examples in atomic physics 
where the system exhibits {\it gain} and $\text{Im}\,n(\omega)$ is negative,
see ref.\cite{Shore:2007um}. See also Sections 6.4 and 7.}

In curved spacetime, however, the assumptions leading to the second 
\eqref{ank} and third \eqref{anl} forms of the Kramers-Kronig relation 
need to be reassessed. We still maintain the causality conditions,
that $\Bn(u;\omega)$ is analytic in the upper-half $\omega$-plane and
bounded at infinity, so the primitive identity \eqref{ani} is always
satisfied. 

Now, in our case, because of the cuts along the real $\omega$-axis
and the definition of the physical sheet, the r.h.s.~of \eqref{anj}
actually involves the function $\Bn_+(u;\omega)$. Then, if we are in a
special case where we have translation invariance along the geodesic,
so that the VVM matrix $\BDelta(u,u')$ is a function only of $(u-u')$,
we have from \eqref{ang} that
\EQ{
\Bn_-(\omega) ~~=~~ \Bn_+(-\omega) \ ,
\label{anm}
}
taking into account the extra $1/\omega$ factor in front of the integral
\eqref{ana} for the refractive index in terms of $\BcalF(z)$.
The r.h.s.~of \eqref{anj} therefore involves
\SP{
\Bn_+(\omega + i\epsilon) - \Bn_+(-\omega + i\epsilon) ~~&=~~
\Bn_+(\omega + i\epsilon) - \Bn_-(\omega - i\epsilon) \\
&=~~ \text{Disc}~ \Bn(\omega) \ .
\label{ann}
}
So translation invariance in $u$ would imply that the second form 
\eqref{ank} of the Kramers-Kronig relation holds even in curved
spacetime.

However, as we shall see in a number of examples, it appears that
real analyticity of $\Bn(u;\omega)$ is lost for QED in curved
spacetime. This stems from the need to define the physical sheet
for $\Bn(u;\omega)$ as in Fig.~\ref{p6} in terms of both 
$\Bn_+(u;\omega)$ and $\Bn_-(u;\omega)$, with the cuts in the
complex $\omega$-plane originating directly from the geometry of 
geodesic deviation and the VVM matrix. So the third form \eqref{anl}
of the Kramers-Kronig relation does {\it not} hold in curved spacetime.
Notice, however, that this does not imply there is anything wrong
with causality or microcausality.

\subsection{Causality and the refractive index}

In this section, we show that the two conditions on the refractive index
assumed in the derivation of the Kramers-Kronig relation, {\it viz}~
(i)~ $\Bn(u;\omega)$ is analytic in the upper-half $\omega$-plane, and 
~(ii)~ $\Bn(u;\omega) \rightarrow\Bone$ for large $|\omega|$, are necessary
conditions for causality. The first is a consequence of requiring that the
refractive index at $x(u)$ only depends on influences in the past light cone
of $x(u)$; the second imposes the condition that the wavefront velocity
(which has been identified in previous work as 
the relevant speed of light for causality) is $c$. 

There are many ways to see how the connection between analyticity and causality
arises, both at the level of the refractive index and more generally in the
construction of Green functions obeying micro-causality. The essential 
technical feature is the theorem that the Fourier transform $\tilde f(z)$ 
of a function $f(t)$ which vanishes for $t > 0$ is analytic in the upper-half
complex $z$-plane. This argument naturally appears in some guise in all
the discussions linking causality with analyticity.

\begin{figure}[ht] 
\centerline{\includegraphics[width=2.5in]{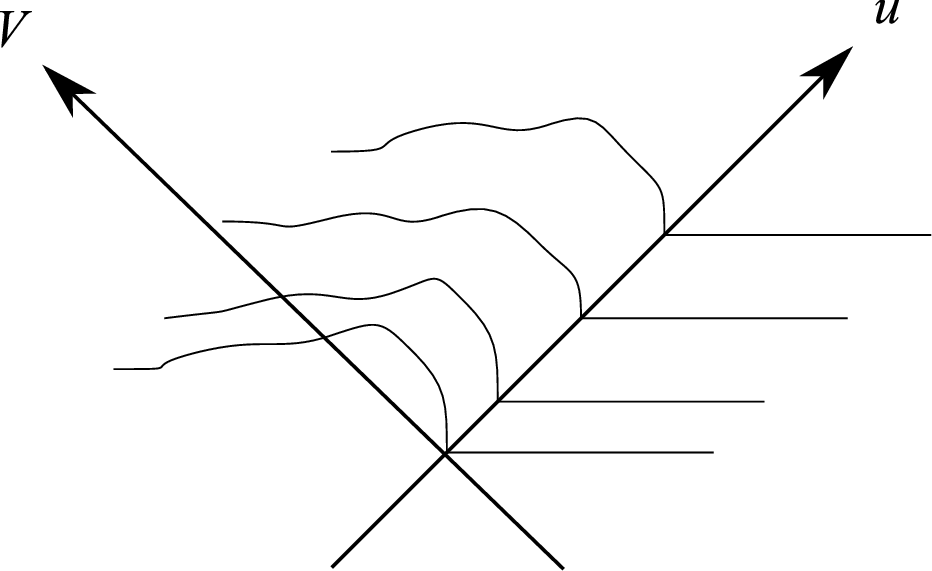}}
\caption{\footnotesize A sharp-fronted wavepacket propagating with
  respect to the light cone.}\label{p22}
\end{figure}

An illuminating illustration is to consider the propagation of a sharp-fronted
wave packet. Consider the one-loop corrected modes  \eqref{dds}. 
Suppose that in the distant past,
$u\to-\infty$, we build a sharp-fronted wave-packet propagating in the
$u$ direction, by taking a Fourier Transform of the modes 
\eqref{dds}:\footnote{Implicitly we
  have been assuming that in the distant past the curvature is turned off
  and the refractive index is asymptotically $\Bone$. Later we will be 
  able to remove this restriction when we discuss Green functions.} 
\EQ{
A^\mu=\int_{-\infty}^\infty d\omega \,{\cal Z}(\omega)\varepsilon^\mu_{(j)}(u)
e^{-i\omega (V-\vartheta_{ij}(u;\omega))}\ ,
\label{wpk}
}
with
\EQ{
\Bvartheta(u;\omega)=\tfrac12
\int_{-\infty}^u du'\,\big[\Bn(u';\omega)-\Bone\big]
}
In the limit, $u\to-\infty$, the refractive index must approach $\Bone$
and so the
condition that \eqref{wpk} be sharp-fronted, that is vanishing for $V<0$,
is that ${\cal Z}(\omega)$ is analytic in the upper-half plane. This
is because when $V<0$ we can deform the integration contour in
\eqref{wpk} from the real axis into the upper-half plane and out to
the semi-circle at infinity on which the integrand
vanishes.\footnote{In order for the wavepacket to be properly defined
  in the presence of the one-loop correction, the integration contour
  must lie just above the real axis to avoid any non-analyticities of
  $\Bn(u;\omega)$ on the real axis.} This
contour deformation argument
manifests the usual link between causality and
analyticity. It follows that at 
finite $u$, the wavepacket will remain sharp-fronted and vanishing
outside the light cone $V<0$
provided that (i) the refractive index 
$\Bn(u;\omega)$, is an analytic function of $\omega$ 
in the upper-half plane; and (ii) $\Bn(u;\omega)\to\Bone$ 
as $|\omega|\to\infty$.  If both these
conditions are satisfied then the sharp-fronted disturbance will
propagate causally.\footnote{Notice that causality might be respected
  if $\Bn(u;\omega)$ approached a constant for large $|\omega|$ but with
  respect to a modified light cone given by
  $V=\vartheta(u;\infty)$. This would rely on the space being suitably
  ``causally stable'' 
   \cite{HawkingEllis,Liberati:2001sd,Shore:2003jx}. 
  The examples that we find do not have
  this property and so we will not pursue this idea.}

To see how these conditions are realized here, consider the expression
\eqref{qya},\eqref{qybb} for the refractive index in terms of an 
integral over $t$ of a function of the VVM matrix $\BDelta(u,u-t)$. Causality requires this to depend only on the part of the geodesic to the past of $x(u)$,
which is guaranteed by the integral being only over $t>0$. But 
since $\BcalF_+(u;z)$ then has the form of a Fourier transform of a 
function which vanishes for $t<0$, it follows from the above theorem that 
$\BcalF_+(u;z)$ is analytic in the lower-half $z$-plane. In turn, this implies analyticity of $\Bn(u;\omega)$ in the upper-half $\omega$-plane.

For the second condition (ii), note that the wavefront velocity is the 
high-frequency limit of the phase velocity.\footnote{This is proved 
in ref.\cite{Leontovich} (see also \cite{Shore:2003jx}) for a very 
general class of wave equation; this proof may not, however, be sufficiently general to cover the full vacuum-polarization induced wave equation \eqref{erg}.} 
For large $|\omega|$, we have 
\EQ{
\Bn(u;\omega)=\Bone-\frac{\alpha}{12\pi\omega}\BcalF(u;0)+\cdots
\label{oot}
}
So a sufficient condition for the second requirement to be
satisfied is that the integral $\BcalF(u;0)$ is finite.\footnote{It 
  might have been possible
  for ${\cal F}(u;z)$ to have a simple pole at $z=0$, in which case
  the high frequency phase velocity is finite but different from
  $c$. But as we have already mentioned this does not occur.}
In particular, this requires that the integral
\SP{
\BcalF_+(u;0)&=i\int_0^{\infty-i\epsilon}\frac{dt}{t^2}\,
\left[\BDelta\big(u,u-t\big)\sqrt{\det\BDelta\big(u,u-t\big)}-\Bone
\right]\ ,
\label{qyb}
}
is convergent. This is actually guaranteed 
by the fact that, as we have already mentioned, implicitly we are
assuming that the space becomes flat in the infinite past and future in order
that the one-loop corrected photon modes be defined consistently in the
whole of spacetime. In that case, for large $t$, $\BDelta(u,u-t)\to1$
as $t\to\infty$.\footnote{We can also discuss spaces which do not
  become flat in the infinite past and future. In that case, the relevant
  problem to consider is an initial value problem and this inevitably
  involves the Green functions, a topic that we turn to in Section 9.}

\vfill\eject

\subsection{Dispersion and $\text{Im}\,\Bn(\omega)$}

In general, the refractive index is a complex quantity. While the real
part determines the local phase velocity at a point, the imaginary
part describes dispersion. To see this we note that the probability
density of the photon wavefunction, with the one-loop correction
include, is
\EQ{
A_{(i)\mu}(u)A_{(j)}^\mu(u)^\dagger=g(u)^{-1/2}
\exp\Bigl[-\omega\int^u_{-\infty}du'\,\text{Im}\,n_{ij}(u';\omega)
\Bigr] \ .
}
The pre-factor here is just the volume effect one would expect in
curved spacetime. The exponential term, on the contrary, determines the dispersive effect of spacetime. In general, we would expect that the eigenvalues of the imaginary part of the refractive index would be $\geq0$
so that spacetime would act like an ordinary dispersive medium with
the total number of photons being depleted as they propagate by
conversion into real $e^+e^-$ pairs. 

However, when we look at some of the examples that we have considered,
in particular the Ricci-flat symmetric plane wave whose refractive
index is plotted numerically in Fig.~\ref{p14} and the weak
gravitational wave in Fig.~\ref{p16}, we see that the imaginary part 
of the refractive index can be negative. Indeed, in the gravitational 
wave example it oscillates sinusoidally with the frequency of the background wave. Apparently, in these cases, spacetime acts as an amplifying medium for photon propagation.

In many ways, this effect is similar to that studied in some atomic
physics examples in \cite{Shore:2007um}. It was shown there that for certain three-state $\Lambda$-systems interacting with coupling and
probe lasers, the refractive index can be arranged to be of the form
\EQ{
n(\omega) ~=~ 1 - {A\over {\omega - \omega_0 + i\gamma}} \ ,
\label{lambda}
}
where $\omega_0$ is a characteristic frequency of the coupled laser-atom system. In the usual dispersive case, we would have $A > 0$ so that
$\text{Re}\,n(0) > 1$ and the low-frequency propagation is subluminal
while $\text{Im}\,n(\omega) > 0$. However, in Raman gain systems we can
arrange to have $A < 0$, resulting in a superluminal $\text{Re}\,n(0)<1$
with $\text{Im}\,n(\omega) < 0$. The negative imaginary part indicates 
that the probe laser is amplified (taking energy from the coupling laser),
with the system acting as an optical medium exhibiting gain.

For our purposes, the important point is that in this model, 
$n(\omega)$ is characterized by a simple pole at 
$\omega = \omega_0 - i\gamma$. This must be in the negative imaginary 
half-plane to be consistent with causality. It then follows that the existence of $\text{Im}\,n(\omega) < 0$ is necessarily linked
to superluminal low-frequency propagation. In the examples below, 
we also find that the occurrence of an imaginary part for the refractive index is correlated with the occurrence of singularities, 
in this case branch points, in $n(\omega)$ off the real axis but in the 
causally-safe half-plane. In turn, the location of these singularities 
is intimately related to the location of singularities of the VVM matrix
in the complex $u$-plane, with polarizations exhibiting 
$\text{Im}\,n(\omega) < 0$ and a superluminal $\text{Re}\,n(0) < 1$ corresponding to the diverging direction of the null geodesic congruence.
However, we should be cautious about over-interpreting our results in this way, since the actual QFT results for the refractive index are 
significantly more complicated than \eqref{lambda}.

It is also important to recognize that this amplification occurs for
photons of high frequency $\frac{\omega^2 R}{m^4} \sim 1$, 
{\it i.e.} $\omega \sim \frac{1}{\lambda_c} \frac{L}{\lambda_c}$ 
where $L$ is the curvature scale. It is not a long-range, infra-red effect
with photon wavelengths comparable to the curvature, 
$\omega \sim \frac{1}{L}$. Rather, the effect seems to be a kind of
emission of photons induced by the interaction of the
incident wave with quantum loops in the curved spacetime background.
However, the details of this mechanism remain to be fully understood.

\section{Example 1: symmetric plane waves}

To illustrate these general results, we now consider some simple examples.
The simplest case is when the background has a Penrose limit which
is a symmetric plane wave. In this case, the matrix functions
$R_{uiuj}(u)$ are independent of $u$. They can immediately be 
diagonalized, $R_{uiuj}=\sigma_i^2\delta_{ij}$, with the $\sigma_i$
constant. The metric in Brinkmann coordinates therefore takes the form
\EQ{
ds^2=-2du\,dv-\sigma_i^2y^i\,y^i\,du^2+dy^i\,dy^i \ .
\label{exa}
}
The non-vaishing component of the Ricci tensor is $R_{uu} = 
\sigma_1^2+\sigma_2^2$.

The VVM matrix can be determined by solving the Jacobi equations 
\eqref{hyw} and \eqref{hh}. This gives
$y^i(u)=c_1\cos(\sigma_iu+c_2)$ and implementing the appropriate boundary
condition in \eqref{hh} selects 
\EQ{
A_{ij}(u,u')=\delta_{ij}\frac{\sin\sigma_i(u-u')}{\sigma_i} \ .
\label{exb}
}
Hence 
\EQ{
\Delta_{ij}(u,u')=\delta_{ij}\frac{\sigma_i(u-u')}
{\sin\sigma_i(u-u')}\ . 
\label{exc}
}
The matrix of refractive indices is independent of $u$ and 
diagonal with elements given by eqs.~\eqref{ana} -- \eqref{and}
with
\EQ{
\BcalF_+(z)=\delta_{ij}
\int_0^{\infty-i\epsilon}\frac{dt}{t^2}\,ie^{-izt}\left[\frac{\sigma_it}{\sin\sigma_it}\prod_{j=1}^2
\sqrt{\frac{\sigma_jt}{\sin\sigma_jt}}-1\right]\ ,
\label{wqq}
}
and similarly for $\BcalF_-(z)$.\footnote{Notice that this
  result is slightly different from that quoted in
  \cite{Hollowood:2007ku,Hollowood:2007kt}. The
  difference is because of the way the overall position of the loop was
  fixed. In \cite{Hollowood:2007ku,Hollowood:2007kt} 
the centre of the loops were fixed by
  hand to be at the origin. 
In the present work we have not needed to fix the overall
  position of the loops since this is done automatically because the
  loops are pinned at $x(0)$ to go through $x$, the origin in Rosen
  coordinates. It turns out that there is a non-trivial Jacobian
  between these prescriptions.}
Notice that in general the integrand has branch-point singularities on
the real $t$ axis since at least one of the $\sigma_j$ is real: these
are the conjugate point singularities. When
one of the $\sigma_j$ is imaginary there are also branch points on the
imaginary axis. 

\subsection{The conformally flat symmetric plane wave}

Consider first the conformally flat symmetric plane wave, with
$\sigma_1=\sigma_2=\sigma$. In this case, both polarizations propagate
with the same phase velocity and refractive index -- there is no
birefringence. We can therefore set $\BcalF(z)= \Bone{\cal F}(z)$.
Eq.\eqref{wqq} simplifies to
\EQ{
{\cal F}_+(z)=\int_0^{\infty-i\epsilon}\frac{dt}{t^2}\,ie^{-izt}
\left[\Big(\frac{\sigma t}{\sin\sigma t}\Big)^2-1\right]\ .
\label{qyc}
}
The integral can be evaluated by first rotating the contour to the
negative imaginary axis and
then by direct evaluation, giving a closed-form expression
in terms of di-gamma functions:
\EQ{
{\cal F}_+(z) = z\log\tfrac{z}{2\sigma} - z\,\psi(1+\tfrac
z{2\sigma})+\sigma\ .
\label{exd}
}
As expected, this is a branched function due to the presence of the 
logarithm. 

The corresponding function ${\cal F}_-(z)$ defined as
\EQ{
{\cal F}_-(z)=\int_0^{-\infty+i\epsilon}\frac{dt}{t^2}\,ie^{-izt}
\left[\Big(\frac{\sigma t}{\sin\sigma t}\Big)^2-1\right]\ .
\label{exe}
}
is given explicitly by
\EQ{
{\cal F}_-(z) = z\log(-\tfrac{z}{2\sigma}) - z\,\psi(1-\tfrac
z{2\sigma})-\sigma\ .
\label{exf}
}
It satisfies
\EQ{
{\cal F}_-(z) ~=~ - {\cal F}_+(-z) \ ,
\label{exg}
}
by virtue of the translation invariance of the symmetric plane wave
metric, which guarantees that the VVM matrix \eqref{exc} is a function
only of $(u-u')$ and the factor $\big[\ldots\big]$ in the integrand
of \eqref{qyc} and \eqref{exe} is an even function of $t$.

\begin{figure}[ht] 
\centerline{\includegraphics[width=2.5in]{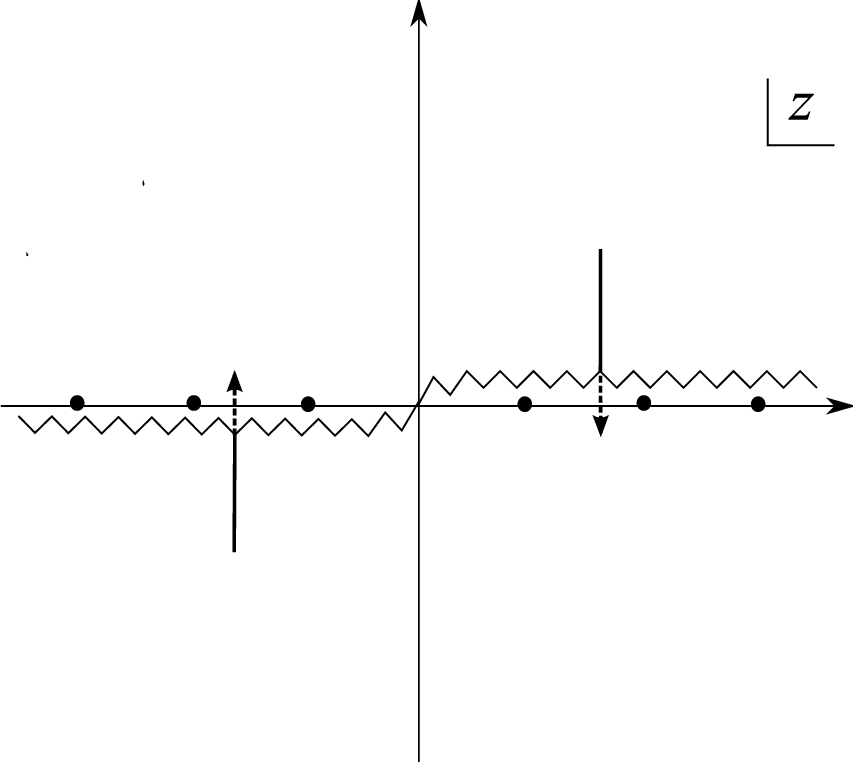}}
\caption{\footnotesize The physical sheet for 
${\cal F}(z)$ showing the branch cuts and the simple poles which lie
on other sheets accessed by moving through the cuts in the direction
of the arrows.}\label{p4}
\end{figure}

The physical sheet is given by the cut $z$-plane with the physical 
${\cal F}(z)$ defined as the analytic continuation of ${\cal F}_+(z)$
into the lower-half plane and of ${\cal F}_-(z)$ into the upper-half
plane, that is (see Fig.~\ref{p4}):
\SP{
{\cal F}(z) ~~=~~\begin{cases}
{\cal F}_+(z) = z\log\tfrac{z}{2\sigma} - z\,\psi(1+\tfrac
z{2\sigma})+\sigma \ ,~~~~~-\pi < {\rm arg~}z \le 0 \ , \\
{\cal F}_-(z) = z\log(-\tfrac{z}{2\sigma}) - z\,\psi(1-\tfrac
z{2\sigma})-\sigma \ ,~~~~~~~0 < {\rm arg~}z \le \pi \ . \end{cases}
\label{exh}
}

\vskip0.2cm
Before considering the discontinuities and the Kramers-Kronig relation, 
notice that in addition to the cuts from the logarithm, ${\cal F}_+(z)$
also has simple poles on the negative real axis at $z = -2p\sigma$, ~
$p = 1,2,\ldots$ from the di-gamma functions.\footnote{$\psi(x)$ has simple 
poles at  $0,-1,-2,\ldots$ with residue $-1$. The di-gamma function also 
satisfies the following identities, to be used later:
$$
\psi(1+z) = \psi(z) + \tfrac1z \ , ~~~~~~~~
\psi(1-z) = \psi(z) + \pi \cot\pi z \ . 
$$} Similarly, ${\cal F}_-(z)$
has poles on the positive real $z$-axis. These poles are not on the
physical sheet, as defined above, so do not directly affect the physical
refractive index. Nevertheless, they encode useful information 
about the functions ${\cal F}_+(z)$ and ${\cal F}_-(z)$ and provide an
alternative method of computing the physical discontinuities.

The full analytic structure of ${\cal F}_+(z)$ can be understood as 
follows. First, introduce a cut-off $\delta$ to regularize the lower
limit of the $t$ integral. It is then useful to consider the integral
as the sum of two pieces. The first term is 
\SP{
{\cal F}_+^{(1)}(\delta,z)
&=-\int_\delta^{\infty-i\epsilon}\frac{dt}{t^2}\,ie^{-izt}\\ &=-\frac{e^{-i\delta
    z}}{\delta}-iz\, \text{Ei}(-i\delta z)\\ &
=-\frac i\delta+\frac{z}2\log(-z^2\delta^2)+z(\gamma_E-1)\ .
\label{asz}
}
where the limit $\delta\to0$ was taken in the last line.
What is interesting is that this term accounts for the branched nature
of ${\cal F}_+(z)$; indeed as $z\to ze^{2\pi i}$, the exponential
integral function has a logarithmic branch cut and so the
discontinuity of \eqref{asz} is $-2\pi z$. The second term,
\EQ{
{\cal F}_+^{(2)}(\delta,z)=i\sigma^2\int_\delta^{\infty-i\epsilon}
dt\,
\frac{e^{-izt}}{\sin^2\sigma
  t}\ ,
}
only has simple poles which can be manifested by expanding the
denominator in powers of $e^{-2i\sigma t}$:
\EQ{
{\cal F}_+^{(2)}(\delta,z)=-4i\sigma^2\int_\delta^{\infty-i\epsilon}
dt\,
\sum_{p=1}^\infty pe^{-i(z+2p\sigma)t}\ .
}
Performing the $t$ integral
\SP{
{\cal F}_+^{(2)}(\delta,z)&=-4\sigma^2e^{-iz\delta}
\sum_{p=1}^\infty \frac{p}{z+2p\sigma}e^{-2ip\sigma\delta}\\
&=\frac{2\sigma e^{-iz\delta}}{e^{2i\sigma\delta}-1}
+\frac{2\sigma z
  e^{-i(z+2\sigma)\delta}}{z+2\sigma}{}_2F_1\big(1,1+\tfrac z{2\sigma};
2+\tfrac z{2\sigma},e^{-2i\sigma\delta}\big)\\ &
=\frac
i\delta-\frac{z}2\log(-\tfrac{4\delta^2}{\sigma^2})-z(\gamma_E-1)+\sigma-z\,
\psi(1+\tfrac
z{2\sigma})\ .
\label{asx}
}
Summing the two contributions \eqref{asz} and \eqref{asx}, we see that
the divergent terms cancel to leave the finite piece \eqref{exd}.

\vskip0.2cm
Returning to the refractive index, we now perform the integral 
over $\xi$ in 
\eqref{ana} and define the physical $n(\omega)$ on the physical sheet
described in Fig.~\ref{p6} as the analytic function found by continuing
$n_+(\omega)$ into the upper-half plane and $n_-(\omega)$ into the 
lower-half plane, that is  
\SP{
n(\omega) ~~=~~ \begin{cases}
n_+(\omega) \ , ~~~~~~~~~~~~~~~0 \le \text{arg}~\omega < \pi \ , \\
n_-(\omega) \ , ~~~~~~~~~~~~~-\pi \le \text{arg}~\omega < 0 \ .
\end{cases}
\label{exi}
}
The translation invariance property ${\cal F}_-(z) = - {\cal F}_+(-z)$
implies
\EQ{
n_-(\omega) ~~=~~ n_+(-\omega) \ .
\label{exj}
}
Also, note that $\text{Im}\,n(\omega) = 0$ for real $\omega$ in this 
example.  With this definition, we also see that $n(\omega)$ is not
a real analytic function, {\it i.e.}~~ $n(\omega^*) \ne n(\omega)^*$.
This is because of the difference in the functions $n_+(\omega)$ and 
$n_-(\omega)$, and reflects the need for the cuts in the $z$ and 
$\omega$-planes in the definition of ${\cal F}(z)$ and $n(\omega)$. 
This shows very clearly how the geometry, in the form of conjugate points, 
implies an analytic structure for the refractive index which fails to satisfy 
the usual $S$-matrix and dispersion relation assumptions.

Since the symmetric plane wave metric exhibits translation invariance in
$u$, the second form of the Kramers-Kronig relation \eqref{ank} should
hold in this example. We now check this. First, we need the discontinuity
$\text{Disc}~{\cal F}(z)$ on the positive real $z$-axis. Using
standard di-gamma function identities (see footnote 19),
we find\footnote{As a consistency check, we can verify that the sum of
the discontinuities on the positive and negative axes, {\it viz.}
$$
{2 i\pi z \over e^{i\pi z/\sigma} - 1} + {2 i \pi z \over
e^{- i\pi z/\sigma} - 1} = - 2 i \pi z \ ,
$$
which reproduces the discontinuity of the logarithmic cuts in 
${\cal F}(z)$.}
\SP{
\text{Disc}~{\cal F}(z) ~~&=~~{\cal F}_-(z) ~-~ {\cal F}_+(z) \\
&=~~-\pi z \cot{\pi z\over 2\sigma} ~+~ i \pi z \\
&=~~-{2 i\pi z \over e^{i\pi z/\sigma} - 1} \ .
\label{exl}
}
This can also be found by using the contour in \eqref{ane}, 
see Fig.~\ref{newfig2},
to evaluate $\text{Disc}~{\cal F}(z)$ as the sum of the residues of the poles 
on the real $t$-axis in the integrand of \eqref{qyc}.
These are double poles at $n\pi/\sigma$, $n\in{\bf Z}\neq0$.
The fact that the singularities are poles rather than branch points is
due to the fact that for the conformally flat symmetric plane wave
conjugate points are simultaneously conjugate for both
polarizations. The discontinuity in ${\cal F}(z)$ associated to the
series of double poles at $t=n\pi/\sigma$ is then given by \eqref{ane} as
\SP{
\text{Disc}~{\cal F}(z)~&=~2\pi i\sum_n \text{Res}_{n\pi/\sigma}\frac
{ie^{-izt}}{t^2}\left[1-\Big(\frac{\sigma t}{\sin\sigma t}\Big)^2
\right]\\
&=~ -2\pi iz \sum_n e^{-i\pi nz/\sigma} \\
&=~ -{2 i \pi z\over e^{i\pi z/\sigma} - 1} \ .
\label{exm} 
}
reproducing the result above.

Finally, substituting back into the refractive index formula, we can
evaluate the Kramers-Kronig relation:
\SP{
n(0) - n(\infty) ~~&=~~{1\over i\pi}~
\int_0^{\infty}\frac{d\omega}{\omega}\,
\text{Disc}~n(\omega-i\epsilon)  \\
&=~~ {\alpha\over 2 i \pi^2}~\int_0^{\infty} 
\frac{d\omega}{\omega^2}\,
\int_0^1 d\xi\, \xi(1-\xi)\, \text{Disc}\,{\cal F}\Big(
\frac{m^2}{2\omega\xi(1-\xi)}+i\epsilon\Big) \\
&=~~ {\alpha\over i \pi^2 m^2}~\int_0^1 d\xi\, [\xi(1-\xi)]^2\,
\int_0^{\infty} dz~\text{Disc}~{\cal F}(z+i\epsilon) \\
&=~~ -{\alpha\over i\pi^2 m^2} \cdot 
{1\over30}\cdot \int_0^{\infty+i\epsilon} dz~
{2i\pi z\over e^{i\pi z/\sigma} - 1} \\
&=~~ -{\alpha\over i\pi^2 m^2} \cdot 
{1\over30}\cdot\Big[-i \int_0^{\infty} dz~
{2\pi z\over e^{\pi z/\sigma} - 1}-8\pi i\sigma^2\sum_{n=1}^\infty n\Big] \\
&=~~- {\alpha \sigma^2 \over 90\pi m^2} \ . 
\label{exn}
}
where the sum over the residues in the second to last line is evaluated as 
$\zeta(-1)=-\tfrac1{12}$.
The $i\epsilon$ prescriptions here are crucial because
$\text{Disc}\,{\cal F}(z)$ has a set of simple poles and the
integration contour must be defined appropriately. The required definition
follows from a close examination of \eqref{ann}. This shows that 
``$\text{Disc}~n(\omega)$" is in fact 
$n_+(\omega + i\epsilon) - n_-(\omega - i\epsilon)$
and picks up not just the discontinuity across the cut itself but
also a contribution from the hidden poles on the unphysical sheet
for $\text{Re}\,\omega > 0$.

We can check this explicitly. For small $\omega$, we have from 
\eqref{nzero} that
\EQ{
n_{ij}(0)=\delta_{ij}-\frac{\alpha}{360\pi
  m^2}\big(R_{uu}\d_{ij}+2R_{uiuj}\big)~~=~~\delta_{ij}- {\alpha\sigma^2\over
90\pi m^2}  \delta_{ij} \ .
\label{nzero2}
}
For large $|\omega|$, the refractive index is determined
by expanding \eqref{qyc} for small $z$. In particular, in the limit as
$|\omega|\to\infty$, we have \eqref{oot}
\EQ{
n(\omega)=\Big[1-\frac{\alpha}{12\pi\omega}{\cal
  F}(0)+\cdots\Big]\ ,
}
plus less singular terms, where
\EQ{
{\cal F}(0)=
i\int_0^{\infty-i\epsilon}\frac{dt}{t^2}\,
\left[\left(\frac{\sigma t}{\sin\sigma t}\right)^2-1
\right]
=-\int_0^{\infty}\frac{dt}{t^2}\,
\left[\left(\frac{\sigma t}{\sinh\sigma t}\right)^2-1
\right]=\sigma\ . 
}
where we have rotated the contour $t\to-it$. So we verify that
the high-frequency limit of the refractive index is indeed $\delta_{ij}$
as expected, corresponding to a wavefront velocities equal to $c$,
and the Kramers-Kronig identity holds (despite the absence of
a non-vanishing $\text{Im}\,n(\omega)$) by virtue of the contribution
from $\text{Disc}~n(\omega)$ across the cut in the complex $\omega$-plane.

\begin{figure}[ht] 
\centerline{\includegraphics[width=3in]{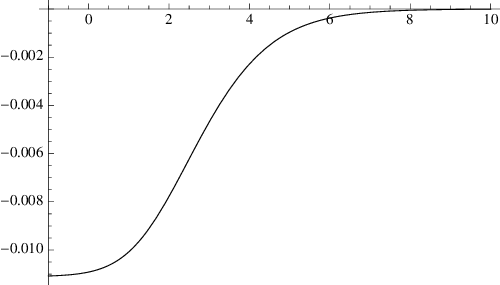}}
\caption{\footnotesize The refractive index $n(\omega)-1$ for the 
conformally flat symmetric plane wave in units of $
\alpha\sigma^2/(\pi m^2)$, as a
  function of $\log\omega\sigma/m^2$.} 
\label{p13}
\end{figure}
The complete frequency dependence of the refractive index can be found
by numerical evaluation of \eqref{qyc} and the result is shown in 
Fig.~\ref{p13}. Notice that because of the sign difference in the one-loop 
coefficients in \eqref{nzero} and \eqref{nzerospin} for scalars and
spinors, the corresponding result for spinor QED is the opposite of
this, {\it viz.}~a superluminal low-frequency phase velocity falling 
monotonically to $c$ in the high-frequency limit.

\subsection{The general symmetric plane wave}

We now extend this analysis to the general symmetric plane
wave \eqref{wqq}. Writing 
${\cal F}_{ij}(z)={\cal F}_j(z)\delta_{ij}$, we have
\EQ{
{\cal F}_{1}(z)=
\int_0^{\infty-i\epsilon}dt\,ie^{-izt}\left[
\frac{\sigma_1^{3/2}\sigma_2^{1/2}}{\sin^{3/2}\sigma_1t\sin^{1/2}
\sigma_2t}-\frac1{t^2}\right]
}
and similarly for ${\cal F}_{2}(z)$ with 
$\sigma_1\leftrightarrow\sigma_2$. 
With no loss of generality we can assume $\sigma_1$ is real, 
while $\sigma_2$ can be real or imaginary with $\sigma_1>|\sigma_2|$ 
(and we shall assume that $\sigma_1$ is positive and 
$\text{arg}\,\sigma_2$ is either 0 or $\tfrac\pi2$). 
To define a physical sheet, we continue  ${\cal F}_{+j}(z)
\equiv{\cal F}_{j}(z)$ into the
lower-half plane (including the positive real axis) and glue it to
${\cal F}_{-j}(z)=-{\cal F}_{+j}(-z)$ in the upper-half plane, just as in
the conformally flat example. 

\begin{figure}[ht]
\centerline{(a)\includegraphics[width=2.5in]{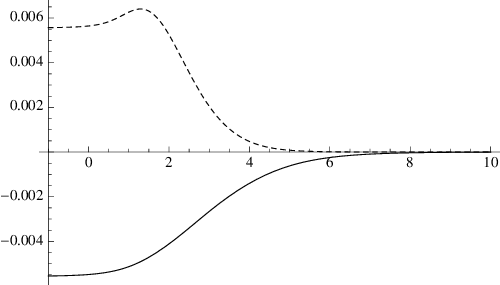}
\hspace{0.2cm}(b)\includegraphics[width=2.5in]{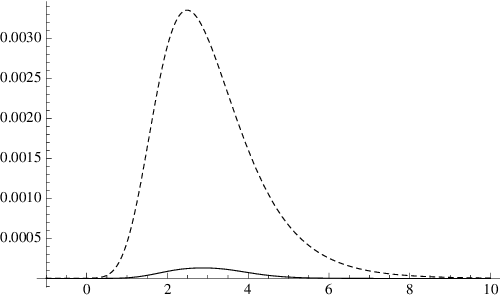}}
\caption{\footnotesize (a) 
$\text{Re}\,n_i(\omega)-1$ and (b) $\text{Im}\,n_i(\omega)$ 
($i=1$ continuous, $i=2$ dashed) in
  units of $\alpha\sigma^2/(\pi m^2)$, as a
  function of $\log\omega\sigma/m^2$ for the Ricci flat
  symmetric plane wave.}\label{p14}
\end{figure}

To demonstrate the new physics arising here, we numerically calculate 
the refractive index for the Ricci flat symmetric plane wave metric, 
$\sigma_1=\sigma$ and $\sigma_2=i\sigma$. 
This displays gravitational birefringence, in
that the two polarizations move with different refractive indices. 
Moreover, in this case, the refractive index also develops an
imaginary part, which would not be seen in the low-frequency 
expansion based on the effective action.  The real and imaginary parts of the refractive indices are plotted in Fig.~\ref{p14}. 

Although we do not have a complete expression for the refractive index
in this case in terms of elementary functions, we can still get a very
accurate approximation using analytic techniques. First of all,
we rotate the contour by taking $t\to-it$:
\EQ{
{\cal F}(z)=-
\int_0^{\infty+i\epsilon}dt\,e^{-zt}\left[\frac{\sigma^2}{
\sin^s\sigma t\sinh^{2-s}\sigma t}-\frac1{t^2}\right]\ ,
}
where $s=\tfrac12$ or $\tfrac32$ depending on the polarization. The
integration contour lies on top of the branch points at $t=n\pi/\sigma$,
$n=1,2,\ldots$. Notice that the integrand is real for $0\leq
t\leq\pi$, $2\pi\leq t\leq3\pi$, {\it etc.\/}, and imaginary for
$\pi\leq t\leq2\pi$, $3\pi\leq t\leq 4\pi$, {\it etc\/}. Since the
integrand is falling off exponentially like $e^{-(z+(2-s)\sigma)t}$ we
can approximate the imaginary part by expanding around the first
branch point $t=\pi/\sigma$; first, taking $s=\tfrac12$:
\SP{
\text{Im}\,{\cal F}_{1}(z)&\simeq -2^{3/2}\sigma^{3/2}
\int_{\pi/\sigma}^\infty dt\,\frac{
e^{-(z+3\sigma/2)t}}{(t-\pi/\sigma)^{1/2}}\\
&=-(2e^{-\pi})^{3/2}\sqrt\pi\sigma^{3/2}\frac{e^{-z\pi/\sigma}}{
\sqrt{z+3\sigma/2}}\ ,
}
and then for $s=\tfrac32$:\footnote{The integral here appears to be
  singular at the lower limit. However, in reality the contour jumps
  over the branch point and this regularizes the integral in a way
  which is equivalent to taking $\int_0^\infty dt\,e^{-t}/t^{3/2}=
\Gamma(-1/2)=-2\sqrt{\pi}$.}
\SP{
\text{Im}\,{\cal F}_{2}(z)&\simeq2^{1/2}\sigma^{1/2}
\int_{\pi/\sigma}^\infty dt\,\frac{
e^{-(z+\sigma/2)t}}{(t-\pi/\sigma)^{3/2}}\\
&=-2 (2e^{-\pi})^{1\over2}\sqrt{\pi}\sigma^{1/2}e^{-z\pi/\sigma}
\sqrt{z+\sigma/2}\ .
}

From these expressions, we 
can get a further approximation of the refractive index itself valid
for low frequency, by evaluating the $\xi$ integral around the
saddle-point of the exponential factor $e^{-z\pi/\sigma}=\exp
-\pi m^2/(2\omega\sigma\xi(1-\xi))$ which occurs at $\xi=\tfrac12$. 
This gives the leading low frequency behaviour as
\EQ{
\text{Im}\,n_1(\omega)=\frac{\alpha\sigma^2}{32\pi m^2}
(2e^{-\pi})^{3/2}\sqrt\pi e^{-2\pi m^2/(\omega\sigma)}+\cdots
}
and
\EQ{
\text{Im}\,n_2(\omega)=\frac{\alpha\sigma}{8\pi\omega}
(2 e^{-\pi})^{1/2}\sqrt{\pi}e^{-2\pi m^2/(\omega\sigma)}+\cdots\ ,
}
which accurately reproduces the numerical evaluation in Fig.~\ref{p14}.

To try and understand the origin of this unusual dispersive behaviour, we can follow the same logic as for the conformally flat example to deduce 
the analytic structure of ${\cal F}_{+j}(z)$, since ultimately the sign
of $\text{Im}\,\Bn(\omega)$ is determined by the location of branch points
on the unphysical sheet of $\Bn(\omega)$. 
The idea is to introduce a cut-off
$\delta$ on the lower limit of the integral and consider the
contribution from the two terms in the integrand.
The contribution \eqref{asz} remains the same, whereas the
second contribution is now
\EQ{
{\cal F}^{(2)}_{1}(\delta,z)=\int_\delta^{\infty-i\epsilon}dt\,ie^{-izt}
\frac{\sigma_1^{3/2}\sigma_2^{1/2}}{\sin^{3/2}\sigma_1t\sin^{1/2}
\sigma_2t}\ .
}
We can expand the integrand in terms of $e^{-2i\sigma_1
  t}$ and $e^{-2i\sigma_2^*t}$, which is a convergent expansion along
the integration contour. (Notice that this is the expansion
  which is consistent with our choice of 
$\text{arg}\,\sigma_2$ to be either 0 or $\tfrac\pi2$.)
 Performing the $t$ integral on the terms in the double expansion gives
\EQ{
{\cal F}^{(2)}_{1}(\delta,z)=-4\sigma_1^{3/2}\sigma_2^{1/2}e^{-iz\delta}
\sum_{m,n=0}^\infty \MAT{\tfrac32\\ m}\MAT{\tfrac12\\ n}
\frac{e^{-i(2m+\tfrac32)\sigma_1\delta-i(2n+\tfrac12)\sigma_2^*\delta}}
{z+(2m+\tfrac32)\sigma_1+(2n+\tfrac12)\sigma_2^*}
\ .
}
While we cannot sum this in closed form, we know that apart from a
singular term which cancels that in ${\cal F}_{+j}(z)$ \eqref{asz}, it is
a holomorphic function with simple poles at
\EQ{
z=-(2m+\tfrac32)\sigma_1-(2n+\tfrac12)\sigma_2^*\ ,\qquad m,n\in{\bf Z}\geq0\ .
\label{ppo}
}
In particular, in the case when $\sigma_2$ is imaginary there are
poles in the upper-half plane. A similar story holds for the other
polarization state.

To conclude, ${\cal F}_{+j}(z)$ has a branch point at $z=0$ coming from
the $z\log z$ term in \eqref{asz} along with a set of simple poles
which lie in the region $\text{Re}\,z<0$, $\text{Im}\,z\geq0$. In
particular since they lie in the upper-half of the $z$ plane they give
rise to branch points in the lower-half of the $\omega$ plane (on an
unphysical sheet) and are therefore in the causally safe region. 

The situation for $n_2(\omega)$, where the polarization lies in the direction of diverging geodesics in the null congruence, is therefore 
quite similar to the simple single-pole refractive index model discussed in Section 6.4 in that we find a low-frequency superluminal phase velocity together with branch points in the negative imaginary half $\omega$-plane.  
The resulting $\text{Im}\,n_2(\omega) < 0$ implies an amplification of
photon propagation through this background spacetime, centred on a 
characteristic frequency of order $\omega^2 R /m^4 \sim 1$.
In contrast, $n_1(\omega)$, where the polarization lies in the direction
of converging geodesics, shows similar behaviour to the conformally flat
example, with a subluminal phase velocity and only a very small,
though still negative, imaginary part $\text{Im}\,n_1(\omega) < 0$.

\section{Example 2: weak gravitational wave}

We now consider an example of a time-dependent background, the
weak gravitational wave, which displays gravitational birefringence 
and dispersion with both positive and negative imaginary parts for
the refractive index.

The spacetime metric for a weak gravitational wave takes the following
form in Rosen coordinates:
\EQ{
ds^2=-2du\,dV+(1+\epsilon \cos\nu u)dY^{1}\,dY^1+\big(1-\epsilon\cos\nu 
u\big)dY^2\,dY^2\ .
}
Here, and in the following, $\epsilon$ is small and we work to linear
order. The transformation to Brinkmann coordinates is achieved via the 
zweibein
\EQ{
E_a{}^j(u)=\delta_a^i\big(1-(-1)^j\frac\epsilon2\cos\nu u\big)\ ,
}
to give
\EQ{
ds^2=-2du\,dv+(-1)^j\frac{\epsilon\nu^2}2\cos\nu u \,y^j y^j\,du^2
+dy^j\,dy^j\ .
}
The equation for the Jacobi fields is
\EQ{
\frac{d^2y^{\sst(\pm)}(u)}{du^2}=\mp\frac{\epsilon\nu^2}2\cos\nu u\, y^j(u)\ ,
}
with $\pm$ corresponding to $j=1$ and $j=2$, respectively.
This can easily be solved perturbatively in $\epsilon$,
with solution to linear order
\EQ{
y^{\sst(\pm)}(u)=c_1+uc_2\pm\frac{\epsilon}2(c_1+uc_2)\cos\nu u
\mp\frac{\epsilon c_2}\nu\sin\nu u\ .
} 
Solving the Jacobi equation \eqref{bct} with the boundary 
condition \eqref{bcs}, we now find the eigenvalues of $\BA$
\SP{
A^{\sst(\pm)}(u,u')&=u-u'\\ &\pm
\frac{\epsilon(u-u')}2\big(\cos\nu u+\cos\nu
u'\big)\mp\frac{\epsilon}{\nu}\big(\sin\nu u-\sin\nu u'\big)\ ,
}
which determines the eigenvalues of the Van-Vleck Morette matrix:
\EQ{
\Delta^{\sst(\pm)}(u,u')=1\mp\frac{\epsilon}2(\cos\nu
u+\cos\nu u')
\pm\frac{\epsilon}{\nu(u-u')}(\sin\nu u-\sin\nu u')\ .
}

The refractive index is given by \eqref{qya} with $\BcalF(u;z)=
\text{diag}{\cal F}^{\sst(\pm)}(u;z)$, where
\SP{
{\cal F}^{\sst(\pm)}(u;z)&=\mp\epsilon
\int_0^{\infty-i\epsilon}\frac{dt}{t^2}\,ie^{-izt}\\ &\times
\Big[\frac12(\cos\nu
u+\cos\nu (u-t))
-\frac1{\nu t}(\sin\nu u-\sin\nu (u-t))\Big]\\
&=\mp\epsilon\big[f_1(z)\cos\nu u+f_2(z)\sin\nu u\big]\ ,
}
where
\SP{
f_1(z)&=\frac{z}{4\nu}\big[2\nu(1+\log
z)+(z-\nu)\log(z-\nu)-(z+\nu)\log(z+\nu)\big]\ ,\\
f_2(z)&=\frac{i}{4\nu}\big[\nu^2+2z^2\log z
+z(\nu-z)\log(z-\nu)-z(z+\nu)\log(z+\nu)\big]\ ,
}

The functions $f_j(z)$ have branch points at $0$, $\infty$ and 
$z=\pm\nu$ and this means that $n_j(u;\omega)$ will have branch points
at 0 $\pm\infty$ and $\pm 2m^2/\nu$. In particular, the branch points
at $\pm 2m^2/\nu$ are points of non-analyticity of the refractive
index. This non-analyticity manifests itself by the fact that
$\text{Im}\,f_1(z)$ and $\text{Re}\,f_2(z)$ are
zero for $z\in{\bf R}>\nu$, while for $z\in{\bf R}<\nu$,
\EQ{
\text{Im}\,f_1(z)=\text{Re}\,f_2(z)=\frac{\pi
  z(\nu-z)}{4\nu}\ .
}

It is then a simple matter to extract the low frequency expansion 
of the refractive index:
\SP{
n^{\sst(\pm)}(u;\omega)&=1\mp\frac{\alpha\epsilon\nu^2}{m^2\pi}
\Big[\frac1{360}+\frac1{6300}\left(\frac{\omega\nu}{m^2}\right)^2+
+\cdots\Big]\cos \nu
u\\
&\pm i
\frac{\alpha\epsilon\nu^2}{m^2\pi}
\Big[\frac1{840}\left(\frac{\omega\nu}{m^2}\right)
+\frac1{10395}\left(\frac{\omega\nu}{m^2}\right)^3+
\cdots\Big]\sin\nu u\ ,
}
while at high frequencies,
\EQ{
n_j(u;\omega)=1\mp i\frac{\alpha\epsilon\nu}{6\pi\omega}\sin \nu
u+\cdots 
}

\begin{figure}[ht] 
\centerline{(a)\includegraphics[width=2.5in]{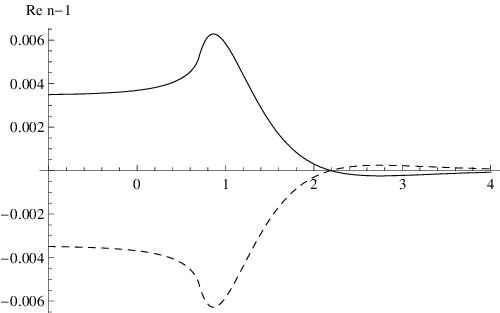}
\hspace{1cm}(b)\includegraphics[width=2.5in]{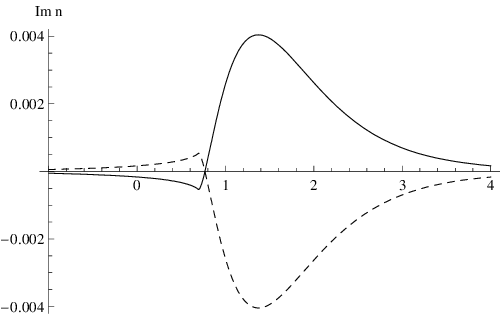}}
\caption{\footnotesize (a) $\text{Re}\,n^{\sst(\pm)}-1$ and (b) 
$\text{Im}\,n^{\sst(\pm)}$ 
for $u=0.2$, $m=\nu=1$
  plotted as a function of $\log\omega$ in units of $\epsilon\alpha$
  for both polarizations. The point of non-analyticity at $\omega=2$
  is quite clear. (Note that the fact that the polarizations
  do not quite give mirror images is an artifact of the numerical
  approximation.}\label{p16}
\end{figure}

The full form of the frequency dependence of the real and imaginary
parts of the refractive index is plotted numerically in Fig.~\ref{p16},
evaluated at a fixed point on the photon trajectory. For the first
polarization, this shows a conventional dispersion for 
$\text{Re}\,n(\omega)$, with a single characteristic frequency 
$\omega^2R/m^4 \sim 1$, together with the corresponding 
$\text{Im}\,n(\omega)>0$. Once more, however, the second polarization
is superluminal at low frequencies and has $\text{Im}\,n(\omega)<0$,
indicating amplification rather than dispersive scattering. The roles
of the two polarizations of course change along the photon trajectory
through the background gravitational wave.

\section{Green Functions}

So far, we have been considering the one-loop correction to photon
modes that come in from past infinity and propagate out to future
infinity. As we have already pointed out, this is, strictly-speaking,
only consistent if the space becomes flat in those limits, otherwise
the one-loop correction to the mode becomes large undermining
perturbation theory. A local way to investigate causality 
involves specifying some
initial data on a Cauchy surface and seeing whether it propagates
causally. This avoids the problem of having modes come in from the
infinite past. Such initial value problems lead to
an investigation of the Green functions. 

\begin{figure}[ht] 
\centerline{\includegraphics[width=2.5in]{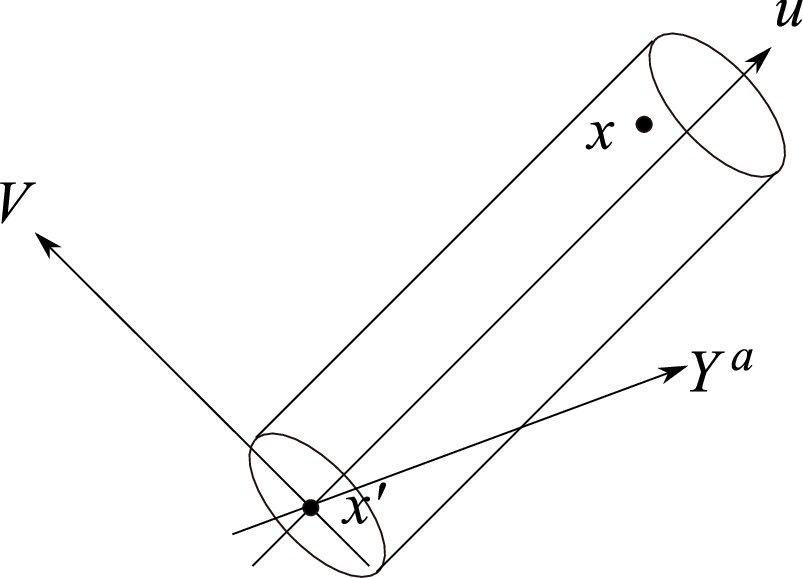}}
\caption{\footnotesize The one-loop correction to the Green
  functions can be calculated for two points $x'=(u',0,0)$ and
  $x=(u,V,Y^a)$ for small $V$ and $Y^a$, so in the neighbourhood of 
the light cone, since the full metric may be approximated by the
  Penrose limit as indicated.}\label{p23}
\end{figure}

For a general spacetime, it is not possible to 
construct the complete Green functions due to the fact that we can only
construct the modes in the eikonal limit $\omega\gg\sqrt R$. However,
if we are interested in the one-loop correction to a Green function
$G(x,x')$, in the neighbourhood of the component of the light cone
with $V=V'=0$ and $Y^a=Y^{\prime a}=0$---and this will teach us about the
one-loop correction to the causal structure---then it is consistent to
replace the full metric by the Penrose limit of the null geodesic
which goes through $(V=0,Y^a=0)$. In this way, we need not work in the
eikonal approximations since the modes \eqref{wkb} are exact in a
plane-wave spacetime. Once we have taken the Penrose limit, then 
it is possible to calculate the one-loop correction to the Green 
functions exactly. (See Fig.\ref{p23}).

In order to construct the Green functions we need a complete set of
on-shell modes. The most immediate problem is that
the general plane wave spacetime does not admit a set of Cauchy
surfaces. However, we will follow \cite{Gibbons:1975jb} and use the null
surfaces $u=\text{const.}$ to define the canonical structure, a choice 
which is sufficient for our purposes. We now search
for the complete set of on-shell modes with respect to the 
inner-product defined on the ersatz Cauchy surfaces:
\EQ{
\big(A,A')=\frac1{i}\int
dV\,d^2Y\,\sqrt{\det\,g(u)}\,\big(A\cdot\partial_V
A^{\prime\dagger}
-A^{\prime\dagger}\cdot\partial_VA\big)\ .
}
The modes $A_{(i)\mu}(x)$, with 
\EQ{
A_{(i)u}(x)=A_{(i)V}(x)=0\ ,\qquad
A_{(i)a}(x)={\cal A}(u)E_{ia}(u)e^{-i\omega V}\ ,
}
that we constructed in Section~3 are clearly on-shell, 
but are not the most general set of modes. 
A complete set of gauge-fixed on-shell modes can be constructed 
by taking a more general eikonal phase and polarization:
\EQ{
A_{(i)\mu}(x)=\varepsilon_{(i)\mu}(x)
\exp-i\big[\omega V+p_aY^a+\psi(u)\big]\ ,
\label{osm}
}
where the eikonal equation \eqref{ngr} determines
\EQ{
\frac{d\psi(u)}{du}=\frac1{2\omega}\big[C^{-1}(u)\big]^{ab}p_ap_b\ ,
}
implying, for later use,
\EQ{
\psi(u)-\psi(u')=\frac{u-u'}
{2\omega}\big[\Delta^{-1}(u,u')\big]^{ab}p_ap_b\ .
}
The gauge-fixed polarization vectors now pick up an additional 
component:
\EQ{
\hat\varepsilon_{(i)a}=E_{ia}(u)\ ,\qquad
\hat\varepsilon_{(i)u}=-\frac{p_aE_i{}^a(u)}{\omega}\ ,
}
while the scalar amplitude remains as in \eqref{sam}.   
Notice, in Brinkmann
coordinates the polarization vector is particularly simple with
$\hat\varepsilon_{(i)j}=\delta_{(i)j}$ and $\hat\varepsilon_{(i)u}$ as
above. The modes are split into 
the positive/negative frequency on-shell modes as
according to whether 
$\omega\gtrless0$ and we will define the 3-momentum $\vec p=(\omega,p_a)$.
One easily finds that the inner-product on the modes is
\EQ{
\big(A_{(i)}({\vec p}),A_{(j)}(\vec p')\big)
=2(2\pi)^3\omega\delta^{(3)}(\vec p-\vec p')\delta_{ij}\
}

All the various propagators can be constructed from these modes. 
We begin by constructing the Wightman functions\footnote{We use the 
  notation of Birrel and Davies \cite{BD}.}
\EQ{
G^\pm_{\mu\nu}(x,x')=\sum_{j=1}^2\int_{\omega\gtrless0} 
\frac{d^3\vec p}{2(2\pi)^3\omega }
  A_{(j)\mu}({\vec p};x)\big(A_{(j)\nu}({\vec p};x')\big)^\dagger
\label{plk}
}
where, as indicated, the
integral over $\omega $ extends over $0$ to $\pm\infty$ for $G^\pm$,
respectively. The $p_a$ integrals are Gaussian and hence easily performed:
\SP{
&\int
d^2p\,\hat\varepsilon_{j\mu}(u)\hat\varepsilon_{j\nu}(u')
\exp\Big[-ip_a(Y-Y')^a-\frac{i(u-u')}{2\omega }
\big[\Delta^{-1}(u,u')\big]^{ab} p_ap_b\Big] 
\\ &=2\pi
\omega\frac{\varpi_{\mu\nu}(x,x')\sqrt{\det\,\Delta_{ab}(u,u')}}
{u-u'}\exp\Big[\frac{i\omega }
{2(u-u')}\Delta_{ab}(u,u')(Y-Y')^a(Y-Y')^b\Big]
\ ,
\label{sls}
}
where $\varpi_{\mu\nu}(x,x')$ has the following non-zero components in Brinkmann coordinates:
\EQ{
\varpi_{ij}(x,x')=\delta_{ij}\
,\qquad\varpi_{uu}(x,x')=\frac{2\text{Tr}\,\BDelta(u,u')}{\omega(u-u')}-
\frac{(y-y')^i(\Delta^2)_{ij}(y-y')^j}{(u-u')^2}\ ,
\label{jus}
} 
We therefore find
\SP{
&G^\pm_{\mu\nu}(x,x')=\frac1{2(2\pi)^2}\frac{\varpi_{\mu\nu}(x,x')
\sqrt{\det\,\BDelta(u,u')}}{u-u'}\\ &
\times\int_{0}^{\pm\infty} d\omega \,\exp\Big[i\omega (V-V') 
+\frac{i\omega }
{2(u-u')}\Delta_{ab}(u,u')(Y-Y')^a(Y-Y')^b\Big]\\
&=\frac1{2(2\pi)^2}\frac{\varpi_{\mu\nu}(x,x')
\sqrt{\det\,\BDelta(u,u')}}{u-u'}
\int_{0}^{\pm\infty}
  d\omega \,\exp\Big[\frac{i\omega \sigma(x,x')}{u-u'}\Big] 
\ ,
\label{tgt}
}
with
\EQ{
\sigma(x,x')=-(u-u')(V-V')+\tfrac12\Delta_{ab}(u,u')(Y-Y')^a(Y-Y')^b\ .
} 

The Feynman propagator $iG_{F\,\mu\nu}(x,x')=\langle
0|T\big(A_\mu(x)A_\nu(x')\big)|0\rangle$, is given by
\EQ{
iG_{F\,\mu\nu}(x,x')=\theta(u-u')G^+_{\mu\nu}(x,x')+\theta(u'-u)
   G^-_{\mu\nu}(x,x')\ .
\label{gyn}
}
However, in the present context we are more interested in the causal
Green functions. In particular, the 
Pauli-Jordan, or Schwinger, function 
\EQ{iG_{\mu\nu}(x,x')=\langle
0|[A_{\mu}(x),A_\nu(x')]|0\rangle
}
 is 
\SP{
&iG_{\mu\nu}(x,x')=G^+_{\mu\nu}(x,x')-G^-_{\mu\nu}(x,x')\\
&=
\frac1{2(2\pi)^2}\frac{\varpi_{\mu\nu}(x,x')\sqrt{\det\,\BDelta(u,u')}}
{u-u'}\int_{-\infty}^\infty
d\omega \,\exp\Big[\frac{i\omega \sigma(x,x')}{u-u'}\Big]\\
&=
\frac1{4\pi}\varpi_{\mu\nu}(x,x')
\sqrt{\det\,\BDelta(u,u')}\delta\big(\sigma(x,x')\big)
\ . 
\label{qww}
}
From this the retarded and advanced Green functions may be extracted
via
\EQ{
G_{R\,\mu\nu}(x,x')=-\theta(u-u')G_{\mu\nu}(x,x')\ ,
\qquad G_{A\,\mu\nu}(x,x')=\theta(u'-u)G_{\mu\nu}(x,x')\ ,\
}

The causal properties of the Pauli-Jordan
function are manifest in the last line of \eqref{qww}. 
$G_{\mu\nu}(x,x')$ has support only if $x$ lies on the forward 
or backward light cone of $x'$. In particular, for 
$G_{R\,\mu\nu}(x,x')$ the support is on the forward light cone 
and for $G_{A\,\mu\nu}(x,x')$ it is on the backward light cone. 
This is precisely what is to be expected for the causality
properties of the Green functions of massless quanta.

The one-loop correction to the Feynman propagator is given by the
usual expression
\EQ{
G^\text{1-loop}_{F\,\mu\nu}(x,x')=-\int d^4\tilde x\,d^4\tilde x'
\,\big(g(\tilde u)g(\tilde u')\big)^{1/2}
G_{F\,\mu}{}^\sigma(x,\tilde x)\Pi^{\text{1-loop}}_{\sigma\rho}(\tilde
x,\tilde x')G_F{}^\rho{}_\mu(\tilde x',x')\ .
\label{iwe}
}
As explained above, the causal properties of a theory are not 
manifested directly in the Feynman propagator, which receives 
contributions both inside and outside the light cone. 
In order to address the causal structure we need to calculate 
the one-loop correction to the Pauli-Jordan function, or
retarded and advanced Green functions. However, these can extracted 
from \eqref{iwe} using \eqref{gyn} and \eqref{qww}. 

The key result is a generalization of the calculation of
Section 4 with the more general on-shell modes \eqref{osm}:
\SP{
&\int d\tilde V\,d^2\tilde Y\,d\tilde V'\,d^2\tilde Y'
\,\sqrt{g(\tilde u)g(\tilde u')}A_{(i)}^\mu(\vec p;\tilde x)^\dagger
\Pi^\text{1-loop}_{\mu\nu}(\tilde x,\tilde x')A_{(j)}^\nu(\vec
p',\tilde x')\\ &
=\frac{2\alpha\omega i}{\pi}(2\pi)^3\delta^{(3)}(\vec p-\vec p') \,
\frac{\Delta_{ij}(\tilde u,\tilde u')\sqrt{\det\BDelta(\tilde u,\tilde
    u')}
}{(\tilde  u-\tilde u')^2} \,
\theta\left(\frac{\tilde u-\tilde u'}{\omega}\right)\\ &\times
\int_0^1d\xi\,\xi(1-\xi)\,\exp\Big[-\frac{im^2(\tilde u-\tilde u')}
{2\omega\xi(1-\xi)}\Big]
+\cdots\ ,
\label{wmm}
}
where the ellipsis indicates additional terms that do not depend on
the curvature, {\it i.e.\/}~are needed to have the correct flat space
limit. 

The strategy for calculating \eqref{iwe} is to write the tree-level 
Feynman propagators in terms of 
$G^\pm$ using \eqref{gyn}. Then we write
$G^\pm$ in terms of the on-shell modes, as in
\eqref{plk}. Once this has been done we can use \eqref{wmm}. The key
point is that \eqref{wmm} conserves the ``momentum'' $\vec p$ and so
the contributions schematically of the form $G^+\Pi G^-$ and $G^-\Pi
G^+$ vanish, leaving the non-vanishing contributions $G^\pm\Pi
G^\pm$ which are immediately identified as the one-loop corrections to
$G^\pm$. Notice that the step functions that are present in
\eqref{gyn} mean that $\tilde u$ and $\tilde u'$ are constrained to be
$u\geq\tilde u\geq\tilde u'\geq u'$ and $u\leq\tilde u\leq \tilde u'\leq u'$,
respectively, for $G^{\text{1-loop}\pm}$. It is then convenient to
change variables from $(\tilde u,\tilde u')$ to $(\tilde u,t)$, where
$t=\tilde u-\tilde u'$. Putting all this together,
we have 
\SP{
G^{\text{1-loop}\pm}_{\mu\nu}(x,x')&=\frac{2i\alpha}{\pi}
\int_{\omega\gtrless0}\frac{d^3p}{2(2\pi)^3\omega}\,\int_{u'}^u d\tilde u\,
\int_0^{u-u'}\frac{dt}{t^2}
\int_0^1d\xi\, \xi(1-\xi)\\ &
\times \Delta_{ij}(\tilde u,\tilde u-t)\sqrt{\det\BDelta(\tilde u,
\tilde u-t)}\\ &\times 
\exp\Big[-\frac{im^2t}{2\omega\xi(1-\xi)}\Big] 
A_{(i)\mu}(\vec p;x)
A_{(j)\nu}(\vec
p;x')^\dagger
+\cdots\ .
\label{wml}
}
which are initially valid for $u\gtrless u'$, respectively, but which
can be extended to all $u$ and $u'$ by analytic continuation.

The $p_a$ integrals in \eqref{wml} 
are identical to \eqref{sls} and so the former
becomes
\SP{
G^{\text{1-loop}\pm}_{ij}(x,x')&=\frac{i\alpha}{(2\pi)^2\pi}\,
\frac{
\sqrt{\det\,\BDelta(u,u')}}{u-u'}
\int_{u'}^ud\tilde u\,\int_0^{u-u'}\frac{dt}{t^2}
\,\int_0^1d\xi\, \xi(1-\xi)\,\\ &
\times
\Delta_{ij}(\tilde u,\tilde u-t)
\sqrt{\det\BDelta(\tilde u,\tilde u-t)}\\ &
\times\int_{0}^{\pm\infty}
  d\omega \,\exp\Big[-\frac{im^2t}{
2\omega\xi(1-\xi)}+\frac{i\omega \sigma(x,x')}{u-u'}\Big]+\cdots\ .
\label{gwn}
}
where we have just displayed the components with spacetime indices
in the two-dimensional polarization subspace.

As happened at tree level, the 
Pauli-Jordan function \eqref{qww} is given by \eqref{gwn} 
by extending the $\omega$
integral from $-\infty$ to $+\infty$:
\SP{
G^{\text{1-loop}}_{ij}(x,x')&=\frac{i\alpha}{(2\pi)^2\pi}\,
\frac{
\sqrt{\det\,\BDelta(u,u')}}{u-u'}
\int_{u'}^ud\tilde u\,\int_0^{u-u'}\frac{dt}{t^2}
\,\int_0^1d\xi\, \xi(1-\xi)\,\\ &
\times
\Delta_{ij}(\tilde u,\tilde u-t)
\sqrt{\det\BDelta(\tilde u,\tilde u-t)}\\ &
\times\int_{-\infty}^{\infty}
  d\omega \,\exp\Big[-\frac{im^2t}{
2\omega\xi(1-\xi)}+\frac{i\omega \sigma(x,x')}{u-u'}\Big]+\cdots\ .
\label{gwntwo}
}
Notice that in the limit $u'\to-\infty$ with fixed $u$, 
the expression can be written in terms of the refractive index. 
In particular, for
$x=(u,V,0,0)$ and $x'=(u',0,0,0)$ in the limit $u'\to-\infty$, the
Pauli-Jordan function (or, since $u>u'$, the retarded propagator) is
\EQ{
G^\text{1-loop}_{R\,ij}(x;x')\thicksim
\int_{-\infty}^u d\tilde u\,\int_{-\infty}^{\infty}
  d\omega \,n_{ij}(\tilde u;\omega)e^{-i\omega V}\ .
\label{gwa}
}
Since $\Bn(u;\omega)$ in analytic in the upper-half plane, the
retarded propagator vanishes when $V<0$, {\it i.e.\/}~outside the light cone. This confirms that even in the presence of the novel dispersion
relations and superluminal phase velocities described here, causality is
maintained with advanced, retarded and Pauli-Jordan Green functions displaying the necessary light-cone support.

A remarkable feature of \eqref{gwntwo} is that we can rewrite it
in a manifestly causal form in terms of the 
Pauli-Jordan function $G(m^2;x,x')$ of a massive scalar particle 
(see eq.\eqref{hyu}):
\SP{
&G^{\text{1-loop}}_{ij}(x,x')=\frac{2\alpha}{\pi}\,
\int_{u'}^u d\tilde u\,\int_0^{u-u'}
\frac{dt}{t^2}\,\int_0^1d\xi\, \xi(1-\xi)\,\\ &
\times
\Delta_{ij}(\tilde u,\tilde u-t)
\sqrt{\det\BDelta(\tilde u,\tilde u-t)} ~ 
G\left(\frac{m^2t}{2\xi(1-\xi)(u-u')};x,x'\right)+\cdots\ .
}
As in previous formulae, the ellipsis represent terms that do not
depend on the curvature. This last expression makes the causal
structure completely manifest. In particular, $G(m^2;x,x')$ has
support only inside, or on, the light cone and so at the one-loop level
the commutator of two photon fields receives contributions from
inside the light cone. However, causality is maintained because the
one-loop correction $G^{\text{1-loop}}_{ij}(x,x')$ still vanishes 
outside the light cone.

\section{Summary and Conclusions}

In this paper, we have analyzed the effect of vacuum polarization in QED
on the propagation of photons through a curved spacetime background.
This problem is of potentially fundamental significance because of the
discovery that quantum loop effects can induce a superluminal phase
velocity, raising the question of how, or whether, this can be
reconciled with causality. We have resolved this issue through an
explicit computation of the full frequency dependence of the
refractive index for QED in curved 
spacetime, showing that the wavefront velocity, which is the speed of light
relevant for causality, is indeed $c$. This is, however, only possible
because a number of generally assumed properties of QFT and $S$-matrix
theory, including the familiar form of the Kramers-Kronig dispersion
relation, do not hold in curved spacetime due to a novel analytic
structure of the Green functions and scattering amplitudes. 

The key insight which makes this analysis possible for general spacetimes
is the realization, inspired by the worldline formalism of QFT, that to
leading order in the curvature, the quantum contributions to photon
propagation are determined by the geometry of geodesic deviation, 
{\it i.e.}~by fluctuations around the null geodesic describing the 
photon's classical trajectory. This geometry is encoded in the Penrose
limit of the original spacetime, simplifying the problem of photon
propagation in general backgrounds to that of their plane wave
limits. 

This allowed us to derive a compact expression for the complete refractive
index $n(\omega)$ of a curved spacetime entirely in terms of the
Van Vleck-Morette matrix for its Penrose limit. In this form, the novel
analytic structure we have discovered becomes manifest. It is related to 
the occurrence of singularities in the VVM matrix corresponding to the
existence of conjugate points in the null geodesic congruence
describing photon propagation. The existence of conjugate points is a
generic property of geodesic congruences, related to the Raychoudhuri
equations and enforced by the null energy condition. This geometrical
origin shows that the type of unconventional analytic structure which
we find here for $n(\omega)$, notably the loss of real analyticity,
will also occur in more general scattering amplitudes in QFT in curved
spacetime. 

The analytic structure of the refractive index and its implications for
causality were described in detail in Section 6. The VVM singularities mean
it is necessary to define $n(\omega)$ on a physical sheet, pasting together
analytic functions on opposite sides of cuts running along the real
$\omega$-axis. This construction banishes dangerous branch-point
singularities to the unphysical sheets, leaving the physical
refractive index with the crucial property of analyticity in the
upper-half complex  
$\omega$-plane required for causality. We also investigated causality
at the level of the Green functions themselves, demonstrating
explicitly how the retarded, advanced and commutator functions have
support on the relevant parts of the fundamental light cone. 

The general theory was illustrated in two examples -- the symmetric plane
wave and the weak gravitational wave background. As well as demonstrating
explicitly how the analytic structure of the refractive index arises and
how the Kramers-Kronig dispersion relation is realized in a way compatible
with causality, these examples introduced a further surprise, 
{\it viz.}~the occurrence of negative imaginary parts for the
refractive index. Such behaviour is generally associated with the
amplification of a light wave as it passes through an optical medium
exhibiting gain, rather than the usual dispersive scattering
associated with the production of  $e^+ e^-$ pairs. It appears that 
in this case quantum loop effects in the curved spacetime background 
are responsible for the emission of photons. 

In addition to clarifying the mechanism responsible for inducing
$\text{Im}\,n(\omega) < 0$ and understanding the relation to the
optical theorem, there are many other directions in which this work
can be extended. An immediate task is to generalize the fundamental
result \eqref{refindex} for the refractive index to spinor QED and 
other QFTs. This requires some further technical developments, but the discussion of analyticity is unlikely to change significantly. 
It is also interesting to extend these results to cosmological and black hole spacetimes and study the effect on photon propagation of horizons and singularities. The phase velocity derived from the low-energy effective action already displays special simplifications at a horizon 
\cite{Shore:1995fz,Gibbons:2000xe}, while the Penrose limit of black
hole spacetimes near the singularity exhibits an interesting
universality \cite{Blau:2003dz} that will be inherited by the refractive index.

Finally, it is important to extend this analysis from photon propagation
and two-point Green functions to higher-point scattering amplitudes.
In view of the geometrical origin of the novel analyticity structure
discovered here, it seems inevitable that many of the conventional
analyticity properties of scattering amplitudes, which underlie all
the usual relations of $S$-matrix theory, will also be radically
changed 
in curved spacetime. In particular, a study of the analytic structure
of Planck energy scattering may well have far-reaching implications
not only for QFT in curved spacetime but also for string theory and
perhaps even quantum gravity itself.

\vspace{0.5cm}
\begin{center}
${\sst **************************}$
\end{center}
\vspace{0.5cm}

We would like to thank Asad Naqvi for discussions, and Luis
Alvarez-Gaum\'e and the CERN Theory Division where this work was
initiated. We also acknowledge the support of STFC grant PP/D507407/1.

\end{document}